\documentclass[a4paper,11pt]{article}
\usepackage[T2A]{fontenc}
\usepackage[utf8]{inputenc}
\usepackage[english]{babel}
\usepackage{amsmath}
\usepackage{amsfonts}
\usepackage{amssymb}
\usepackage{wasysym}
\usepackage{amsthm}
\usepackage{tikz}
\usepackage{tikz-cd}
\usetikzlibrary{decorations.markings}
\usetikzlibrary{positioning}
\usepackage{braids}
\usepackage{array}
\usepackage{ytableau}
\usepackage{longtable}
\usepackage{empheq}
\usepackage{multicol}
\usepackage{mathrsfs}
\usepackage{amssymb}
\usepackage{diagbox}
\usepackage{color}
\usepackage{physics}
\usepackage{ytableau}
\definecolor{lgreen}{rgb}{0.9,1,0.8}

\usepackage[most]{tcolorbox}

\usepackage{pdflscape}
\usepackage{array, longtable}
\newcolumntype{C}{>{$}c<{$}}

\usepackage{pb-diagram}
\usepackage{setspace}
\usepackage[hidelinks,colorlinks=true,unicode]{hyperref}
\hypersetup{
	linkcolor=blue,
	citecolor=blue,
	filecolor=magenta,
	urlcolor=blue,
}
\usepackage{url}
\usepackage[left=2.3cm,right=2.3cm,top=2.3cm,bottom=2.3cm,bindingoffset=0cm]{geometry}

\title{\textbf{Macdonald polynomials for super-partitions}
	\author{???}
	\date{}}

\def\be{\begin{eqnarray}}
\def\ee{\end{eqnarray}}
\def\nn{\nonumber}

\begin{document}
\hfill MIPT/TH-13/24

\hfill ITEP/TH-18/24

\hfill IITP/TH-15/24

\vskip 1.5in
\begin{center}
	
	{\bf\Large Macdonald polynomials for super-partitions}
	
	\vskip 0.2in
	\renewcommand{\thefootnote}{\fnsymbol{footnote}}
	{Dmitry Galakhov$^{2,3,4,}$\footnote[2]{e-mail: galakhov@itep.ru},  Alexei Morozov$^{1,2,3,4,}$\footnote[3]{e-mail: morozov@itep.ru} and Nikita Tselousov$^{1,2,4,}$\footnote[4]{e-mail: tselousov.ns@phystech.edu}}
	\vskip 0.2in
	\renewcommand{\thefootnote}{\roman{footnote}}
	{\small{
			\textit{$^1$MIPT, 141701, Dolgoprudny, Russia}
			\vskip 0 cm
			\textit{$^2$NRC “Kurchatov Institute”, 123182, Moscow, Russia}
			\vskip 0 cm
			\textit{$^3$IITP RAS, 127051, Moscow, Russia}
			\vskip 0 cm
			\textit{$^4$ITEP, Moscow, Russia}
	}}
\end{center}

\vskip 0.2in
\baselineskip 16pt

\centerline{ABSTRACT}

\bigskip

{\footnotesize
We introduce generalization of famous Macdonald polynomials for the case of super-Young diagrams that contain half-boxes on the equal footing with full boxes. These super-Macdonald polynomials are polynomials of extended set of variables: usual $p_k$ variables are accompanied by anti-commuting Grassmann variables $\theta_k$. Starting from recently defined super-Schur polynomials and exploiting orthogonality relations with triangular decompositions we are able to fully determine super-Macdonald polynomials. These new polynomials have similar properties to canonical Macdonald polynomials -- they respect two different orderings in the set of (super)-Young diagrams simultaneously.
	
}

\bigskip

\bigskip

	\ytableausetup{boxsize = 0.5em}

\section{Introduction}
In this paper we propose a definition for generalization of Macdonald polynomials \cite{Macdonald} to the case of super-partitions or super-Young diagrams\footnote{Similar diagrams appear independently in work \cite{Noshita:2021dgj}.}. These super-diagrams can be thought as half-integer partitions $\lambda = \left[ \lambda_1, \lambda_2, \lambda_3, \ldots, \lambda_{l(\lambda)}\right]$, $\lambda_k \in \mathbb{N}/2$ with decreasing rule $\lambda_1 \geqslant \lambda_2 \geqslant \lambda_3 \geqslant \ldots \geqslant \lambda_{l(\lambda)}$. Importantly, if neighboring values $\lambda_k$ and $\lambda_{k+1}$ are both half-integer then inequality necessarily becomes strict $\lambda_k > \lambda_{k+1}$. Usual Young diagrams are the subset of super-Young diagrams. We depict super-partitions as the following diagrams:

\begin{figure}[h!]
	\centering
	\begin{tikzpicture}
		\node(B) at (6,0) {$\begin{array}{c}
				\begin{tikzpicture}[scale=0.5]
					\foreach \i/\j in {0/0, 1/0, 2/0, 3/0, 4/0, 5/0, 6/0, 7/0, 0/-1, 1/-1, 2/-1, 3/-1, 4/-1, 5/-1, 6/-1, 0/-2, 1/-2, 2/-2, 3/-2, 4/-2, 5/-2, 0/-3, 1/-3, 2/-3, 3/-3, 0/-4, 1/-4, 2/-4, 0/-5, 0/-6}
					{
						\draw[thick] (\i,\j) -- (\i+1,\j);
					}
					\foreach \i/\j in {0/0, 0/-1, 0/-2, 0/-3, 0/-4, 0/-5, 1/0, 1/-1, 1/-2, 1/-3, 1/-4, 1/-5, 2/0, 2/-1, 2/-2, 2/-3, 3/0, 3/-1, 3/-2, 3/-3, 4/0, 4/-1, 4/-2, 5/0, 5/-1, 6/0, 6/-1, 7/0 }
					{
						\draw[thick] (\i,\j) -- (\i,\j-1);
					}
					\foreach \i/\j in {8/0, 5/-2, 4/-3}
					{
						\draw[thick] (\i,\j) -- (\i-1,\j-1);
					}
				\end{tikzpicture}
			\end{array}$};
		\node[right] at (B.east) {$=\left[ \frac{15}{2}, 6,\frac{9}{2}, \frac{7}{2}, 1, 1 \right]$};
		\node[left] at (B.west) {$\lambda = $};
	\end{tikzpicture}
	\label{fig:crystalYoung}
\end{figure}
On the pictures $\lambda_k$ is represented by the $k$-th row in the super-Young diagram.  Half-integers are represented by rows with half-box in the end. The number of super-Young diagrams $p(n,m)$ with $n$ full boxes and $m$ half-boxes can be extracted from the following generating function:
\begin{equation}
	\prod_{k=1} \frac{1 + y \, x^{k-1}}{1 - x^{k}} = \sum_{n,m=0} p(n,m) \, x^{n} y^{m} = 1 + y + x + 2 x y + 2 x^2 + x y^2 + 4 x^2 y + \ldots
\end{equation}

If we set $y = 0$ the above generating function becomes generating function for number of usual Young diagrams $p(n)$:
\begin{equation}
	\prod_{k=1} \frac{1}{1 - x^{k}} = \sum_{n,m=0} p(n) \, x^{n} = 1 + x + 2 x^2 + 3 x^3 + 5 x^4 + \ldots
\end{equation}

One of the basic but trivial facts in the theory of Schur/Macdonald polynomials is that the number of Young diagrams of given size $n$ is equal to the dimension of homogeneous polynomials of degree $n$ of infinite variables $p_a$, $a \in \mathbb{N} $, taking into account that the degree of $p_a$ is equal to $a$. Simple examples from the small levels that illustrate this fact are collected in the following Table \ref{Boson table}.
\begin{table}[h!]
	\centering
	$\begin{array}{|c|c|c|}
		\hline
		\text{Level} & \text{Young diagrams} & \text{Basis polynomials} \\
		\hline
		1 & \begin{array}{c}
			\begin{tikzpicture}
			\node(a1) at (0,0) {$\begin{array}{c}
						\begin{tikzpicture}[scale=0.3]
							\foreach \i/\j in {0/0, 0/-1}
							{
								\draw[thick] (\i,\j) -- (\i+1,\j);
							}
							\foreach \i/\j in {0/0, 1/0}
							{
								\draw[thick] (\i,\j) -- (\i,\j-1);
							}
							\foreach \i/\j in {}
							{
								\draw[thick] (\i,\j) -- (\i-1,\j-1);
							}
						\end{tikzpicture}
					\end{array}$};
			\end{tikzpicture}
		\end{array} &
		\begin{array}{c}
			p_1
		\end{array} \\
		\hline
		2 &  \begin{array}{c}
			\begin{tikzpicture}
			\node(a1) at (0,0) {$\begin{array}{c}
					\begin{tikzpicture}[scale=0.3]
						\foreach \i/\j in {0/0, 0/-1, 1/0, 1/-1}
						{
							\draw[thick] (\i,\j) -- (\i+1,\j);
						}
						\foreach \i/\j in {0/0, 1/0, 2/0}
						{
							\draw[thick] (\i,\j) -- (\i,\j-1);
						}
						\foreach \i/\j in {}
						{
							\draw[thick] (\i,\j) -- (\i-1,\j-1);
						}
					\end{tikzpicture}
				\end{array}$};
			\node(a2) at (0.8,0) {$\begin{array}{c}
					\begin{tikzpicture}[scale=0.3]
						\foreach \i/\j in {0/0, 0/-1, 0/-2}
						{
							\draw[thick] (\i,\j) -- (\i+1,\j);
						}
						\foreach \i/\j in {0/0, 1/0, 0/-1, 1/-1}
						{
							\draw[thick] (\i,\j) -- (\i,\j-1);
						}
						\foreach \i/\j in {}
						{
							\draw[thick] (\i,\j) -- (\i-1,\j-1);
						}
					\end{tikzpicture}
				\end{array}$};
		\end{tikzpicture}
		\end{array} &
		\begin{array}{c}
			p_2, p_1^2
		\end{array} \\
		\hline
		3 &  \begin{array}{c}
			\begin{tikzpicture}
			\node(a1) at (0,0) {$\begin{array}{c}
					\begin{tikzpicture}[scale=0.3]
						\foreach \i/\j in {0/0, 0/-1, 1/0, 1/-1, 2/0, 2/-1}
						{
							\draw[thick] (\i,\j) -- (\i+1,\j);
						}
						\foreach \i/\j in {0/0, 1/0, 2/0, 3/0}
						{
							\draw[thick] (\i,\j) -- (\i,\j-1);
						}
						\foreach \i/\j in {}
						{
							\draw[thick] (\i,\j) -- (\i-1,\j-1);
						}
					\end{tikzpicture}
				\end{array}$};
			\node(a2) at (1.15,0) {$\begin{array}{c}
					\begin{tikzpicture}[scale=0.3]
						\foreach \i/\j in {0/0, 0/-1, 0/-2, 1/0, 1/-1}
						{
							\draw[thick] (\i,\j) -- (\i+1,\j);
						}
						\foreach \i/\j in {0/0, 1/0, 0/-1, 1/-1, 2/0}
						{
							\draw[thick] (\i,\j) -- (\i,\j-1);
						}
						\foreach \i/\j in {}
						{
							\draw[thick] (\i,\j) -- (\i-1,\j-1);
						}
					\end{tikzpicture}
				\end{array}$};
			\node(a3) at (2,0) {$\begin{array}{c}
					\begin{tikzpicture}[scale=0.3]
						\foreach \i/\j in {0/0, 0/-1, 0/-2, 0/-3}
						{
							\draw[thick] (\i,\j) -- (\i+1,\j);
						}
						\foreach \i/\j in {0/0, 1/0, 0/-1, 1/-1, 0/-2, 1/-2}
						{
							\draw[thick] (\i,\j) -- (\i,\j-1);
						}
						\foreach \i/\j in {}
						{
							\draw[thick] (\i,\j) -- (\i-1,\j-1);
						}
					\end{tikzpicture}
				\end{array}$};
		\end{tikzpicture}
		\end{array} &
		\begin{array}{c}
			p_3, p_2 p_1, p_1^3
		\end{array} \\
		\hline
		4 &
		\begin{array}{c}
			\begin{tikzpicture}
			\node(a1) at (0,0) {$\begin{array}{c}
					\begin{tikzpicture}[scale=0.3]
						\foreach \i/\j in {0/0, 0/-1, 1/0, 1/-1, 2/0, 2/-1, 3/0, 3/-1}
						{
							\draw[thick] (\i,\j) -- (\i+1,\j);
						}
						\foreach \i/\j in {0/0, 1/0, 2/0, 3/0, 4/0}
						{
							\draw[thick] (\i,\j) -- (\i,\j-1);
						}
						\foreach \i/\j in {}
						{
							\draw[thick] (\i,\j) -- (\i-1,\j-1);
						}
					\end{tikzpicture}
				\end{array}$};
			\node(a2) at (1.4,0) {$\begin{array}{c}
					\begin{tikzpicture}[scale=0.3]
						\foreach \i/\j in {0/0, 0/-1, 0/-2, 1/0, 1/-1, 2/0, 2/-1}
						{
							\draw[thick] (\i,\j) -- (\i+1,\j);
						}
						\foreach \i/\j in {0/0, 1/0, 0/-1, 1/-1, 2/0, 3/0}
						{
							\draw[thick] (\i,\j) -- (\i,\j-1);
						}
						\foreach \i/\j in {}
						{
							\draw[thick] (\i,\j) -- (\i-1,\j-1);
						}
					\end{tikzpicture}
				\end{array}$};
			\node(a3) at (2.45,0) {$\begin{array}{c}
					\begin{tikzpicture}[scale=0.3]
						\foreach \i/\j in {0/0, 0/-1, 0/-2, 1/0, 1/-1, 1/-2}
						{
							\draw[thick] (\i,\j) -- (\i+1,\j);
						}
						\foreach \i/\j in {0/0, 1/0, 0/-1, 1/-1, 2/0, 2/-1}
						{
							\draw[thick] (\i,\j) -- (\i,\j-1);
						}
						\foreach \i/\j in {}
						{
							\draw[thick] (\i,\j) -- (\i-1,\j-1);
						}
					\end{tikzpicture}
				\end{array}$};
			\node(a4) at (3.4,0) {$\begin{array}{c}
					\begin{tikzpicture}[scale=0.3]
						\foreach \i/\j in {0/0, 0/-1, 0/-2, 0/-3, 1/0, 1/-1}
						{
							\draw[thick] (\i,\j) -- (\i+1,\j);
						}
						\foreach \i/\j in {0/0, 1/0, 0/-1, 1/-1, 0/-2, 1/-2, 2/0}
						{
							\draw[thick] (\i,\j) -- (\i,\j-1);
						}
						\foreach \i/\j in {}
						{
							\draw[thick] (\i,\j) -- (\i-1,\j-1);
						}
					\end{tikzpicture}
				\end{array}$};
			\node(a5) at (4.2,0) {$\begin{array}{c}
					\begin{tikzpicture}[scale=0.3]
						\foreach \i/\j in {0/0, 0/-1, 0/-2, 0/-3, 0/-4}
						{
							\draw[thick] (\i,\j) -- (\i+1,\j);
						}
						\foreach \i/\j in {0/0, 1/0, 0/-1, 1/-1, 0/-2, 1/-2, 0/-3, 1/-3}
						{
							\draw[thick] (\i,\j) -- (\i,\j-1);
						}
						\foreach \i/\j in {}
						{
							\draw[thick] (\i,\j) -- (\i-1,\j-1);
						}
					\end{tikzpicture}
				\end{array}$};
		\end{tikzpicture}
		\end{array} &
		\begin{array}{c}
			p_4, p_3 p_1, p_2^2, p_2 p_1^2, p_1^4
		\end{array} \\
		\hline
	\end{array}$
	\caption{Usual Young diagrams are in one-to-one correspondence to the homogeneous polynomials in variables $p_a$.}
	\label{Boson table}
\end{table}

Remarkably for the super-diagrams similar statement holds. The number of super-Young diagrams on a given level is equal to the dimension of homogeneous polynomials of extended set of variables $p_a$ and $\theta_a$, $a \in \mathbb{N}$, where degree of variable $\theta_a$ is equal to $a-1/2$. The main subtlety is that $\theta_a$ variables are Grassmann variables obeying anti-commuting relations:
\begin{equation}
	\Big\{ \theta_a, \theta_b \Big\} = \theta_a \theta_b + \theta_b \theta_a = 0
\end{equation}
In particular, squares of these variables vanish
\begin{equation}
	\theta_a^2 = 0
\end{equation}
that strongly restricts the dimension of corresponding spaces. Using various properties of usual Macdonald polynomials as a base point we define a new set of super-Macdonald polynomials that are polynomials in double set of variable $(p_a, \theta_a)$ and are enumerated by super-Young diagrams.

\begin{table}[h!]
	\centering
	$\begin{array}{|c|c|c|}
		\hline
		\text{Level} & \text{Young diagrams} & \text{Basis polynomials} \\
		\hline
		\frac{1}{2} &
		\begin{array}{c} \begin{tikzpicture}
			\node(a1) at (0,0) {$\begin{array}{c}
					\begin{tikzpicture}[scale=0.3]
						\foreach \i/\j in {0/0}
						{
							\draw[thick] (\i,\j) -- (\i+1,\j);
						}
						\foreach \i/\j in {0/0}
						{
							\draw[thick] (\i,\j) -- (\i,\j-1);
						}
						\foreach \i/\j in {1/0}
						{
							\draw[thick] (\i,\j) -- (\i-1,\j-1);
						}
					\end{tikzpicture}
				\end{array}$};
		\end{tikzpicture}
		\end{array} &
		\begin{array}{c}
			\theta_1
		\end{array} \\
		\hline
		1 &
		\begin{array}{c} \begin{tikzpicture}
			\node(a1) at (0,0) {$\begin{array}{c}
					\begin{tikzpicture}[scale=0.3]
						\foreach \i/\j in {0/0, 0/-1}
						{
							\draw[thick] (\i,\j) -- (\i+1,\j);
						}
						\foreach \i/\j in {0/0, 1/0}
						{
							\draw[thick] (\i,\j) -- (\i,\j-1);
						}
						\foreach \i/\j in {}
						{
							\draw[thick] (\i,\j) -- (\i-1,\j-1);
						}
					\end{tikzpicture}
				\end{array}$};
		\end{tikzpicture}
		\end{array} &
		\begin{array}{c}
			p_1
		\end{array} \\
		\hline
		\frac{3}{2} &
		\begin{array}{c}
			\begin{tikzpicture}
			\node(a1) at (0,0) {$\begin{array}{c}
					\begin{tikzpicture}[scale=0.3]
						\foreach \i/\j in {0/0, 0/-1, 1/0}
						{
							\draw[thick] (\i,\j) -- (\i+1,\j);
						}
						\foreach \i/\j in {0/0, 1/0}
						{
							\draw[thick] (\i,\j) -- (\i,\j-1);
						}
						\foreach \i/\j in {2/0}
						{
							\draw[thick] (\i,\j) -- (\i-1,\j-1);
						}
					\end{tikzpicture}
				\end{array}$};
			\node(a2) at (0.8,0) {$\begin{array}{c}
					\begin{tikzpicture}[scale=0.3]
						\foreach \i/\j in {0/0, 0/-1}
						{
							\draw[thick] (\i,\j) -- (\i+1,\j);
						}
						\foreach \i/\j in {0/0, 1/0, 0/-1}
						{
							\draw[thick] (\i,\j) -- (\i,\j-1);
						}
						\foreach \i/\j in {1/-1}
						{
							\draw[thick] (\i,\j) -- (\i-1,\j-1);
						}
					\end{tikzpicture}
				\end{array}$};
		\end{tikzpicture}
		\end{array} &
		\begin{array}{c}
		\theta_2, p_1 \theta_1
		\end{array} \\
		\hline
		2 &
		\begin{array}{c} \begin{tikzpicture}
			\node(a1) at (0,0) {$\begin{array}{c}
					\begin{tikzpicture}[scale=0.3]
						\foreach \i/\j in {0/0, 0/-1, 1/0, 1/-1}
						{
							\draw[thick] (\i,\j) -- (\i+1,\j);
						}
						\foreach \i/\j in {0/0, 1/0, 2/0}
						{
							\draw[thick] (\i,\j) -- (\i,\j-1);
						}
						\foreach \i/\j in {}
						{
							\draw[thick] (\i,\j) -- (\i-1,\j-1);
						}
					\end{tikzpicture}
				\end{array}$};
			\node(a2) at (0.8,0) {$\begin{array}{c}
					\begin{tikzpicture}[scale=0.3]
						\foreach \i/\j in {0/0, 0/-1, 0/-2}
						{
							\draw[thick] (\i,\j) -- (\i+1,\j);
						}
						\foreach \i/\j in {0/0, 1/0, 0/-1, 1/-1}
						{
							\draw[thick] (\i,\j) -- (\i,\j-1);
						}
						\foreach \i/\j in {}
						{
							\draw[thick] (\i,\j) -- (\i-1,\j-1);
						}
					\end{tikzpicture}
				\end{array}$};
			\node(a3) at (1.7,0) {$\begin{array}{c}
					\begin{tikzpicture}[scale=0.3]
						\foreach \i/\j in {0/0, 0/-1, 1/0}
						{
							\draw[thick] (\i,\j) -- (\i+1,\j);
						}
						\foreach \i/\j in {0/0, 1/0, 0/-1}
						{
							\draw[thick] (\i,\j) -- (\i,\j-1);
						}
						\foreach \i/\j in {2/0, 1/-1}
						{
							\draw[thick] (\i,\j) -- (\i-1,\j-1);
						}
					\end{tikzpicture}
				\end{array}$};
		\end{tikzpicture}
		\end{array} &
		\begin{array}{c}
			p_2, p_1^2, \theta_1 \theta_2
		\end{array} \\
		\hline
		\frac{5}{2} &
		\begin{array}{c} \begin{tikzpicture}
			\node(a1) at (0,0) {$\begin{array}{c}
					\begin{tikzpicture}[scale=0.3]
						\foreach \i/\j in {0/0, 0/-1, 1/0, 1/-1, 2/0}
						{
							\draw[thick] (\i,\j) -- (\i+1,\j);
						}
						\foreach \i/\j in {0/0, 1/0, 2/0}
						{
							\draw[thick] (\i,\j) -- (\i,\j-1);
						}
						\foreach \i/\j in {3/0}
						{
							\draw[thick] (\i,\j) -- (\i-1,\j-1);
						}
					\end{tikzpicture}
				\end{array}$};
			\node(a2) at (1.15,0) {$\begin{array}{c}
					\begin{tikzpicture}[scale=0.3]
						\foreach \i/\j in {0/0, 0/-1, 1/0, 1/-1}
						{
							\draw[thick] (\i,\j) -- (\i+1,\j);
						}
						\foreach \i/\j in {0/0, 1/0, 0/-1, 2/0}
						{
							\draw[thick] (\i,\j) -- (\i,\j-1);
						}
						\foreach \i/\j in {1/-1}
						{
							\draw[thick] (\i,\j) -- (\i-1,\j-1);
						}
					\end{tikzpicture}
				\end{array}$};
			\node(a3) at (2.15,0) {$\begin{array}{c}
					\begin{tikzpicture}[scale=0.3]
						\foreach \i/\j in {0/0, 0/-1, 1/0, 0/-2}
						{
							\draw[thick] (\i,\j) -- (\i+1,\j);
						}
						\foreach \i/\j in {0/0, 1/0, 0/-1, 1/-1}
						{
							\draw[thick] (\i,\j) -- (\i,\j-1);
						}
						\foreach \i/\j in {2/0}
						{
							\draw[thick] (\i,\j) -- (\i-1,\j-1);
						}
					\end{tikzpicture}
				\end{array}$};
			\node(a4) at (3,0) {$\begin{array}{c}
					\begin{tikzpicture}[scale=0.3]
						\foreach \i/\j in {0/0, 0/-1, 0/-2}
						{
							\draw[thick] (\i,\j) -- (\i+1,\j);
						}
						\foreach \i/\j in {0/0, 1/0, 0/-1, 1/-1, 0/-2}
						{
							\draw[thick] (\i,\j) -- (\i,\j-1);
						}
						\foreach \i/\j in {1/-2}
						{
							\draw[thick] (\i,\j) -- (\i-1,\j-1);
						}
					\end{tikzpicture}
				\end{array}$};
		\end{tikzpicture}
		\end{array} &
		\begin{array}{c}
			\theta_3, p_2 \theta_1, p_1 \theta_2, p_1^2 \theta_1
		\end{array} \\
		\hline
		3 & \begin{array}{c}
			\begin{tikzpicture}
			\node(a1) at (0,0) {$\begin{array}{c}
					\begin{tikzpicture}[scale=0.3]
						\foreach \i/\j in {0/0, 0/-1, 1/0, 1/-1, 2/0, 2/-1}
						{
							\draw[thick] (\i,\j) -- (\i+1,\j);
						}
						\foreach \i/\j in {0/0, 1/0, 2/0, 3/0}
						{
							\draw[thick] (\i,\j) -- (\i,\j-1);
						}
						\foreach \i/\j in {}
						{
							\draw[thick] (\i,\j) -- (\i-1,\j-1);
						}
					\end{tikzpicture}
				\end{array}$};
			\node(a2) at (1.15,0) {$\begin{array}{c}
					\begin{tikzpicture}[scale=0.3]
						\foreach \i/\j in {0/0, 0/-1, 0/-2, 1/0, 1/-1}
						{
							\draw[thick] (\i,\j) -- (\i+1,\j);
						}
						\foreach \i/\j in {0/0, 1/0, 0/-1, 1/-1, 2/0}
						{
							\draw[thick] (\i,\j) -- (\i,\j-1);
						}
						\foreach \i/\j in {}
						{
							\draw[thick] (\i,\j) -- (\i-1,\j-1);
						}
					\end{tikzpicture}
				\end{array}$};
			\node(a3) at (2,0) {$\begin{array}{c}
					\begin{tikzpicture}[scale=0.3]
						\foreach \i/\j in {0/0, 0/-1, 0/-2, 0/-3}
						{
							\draw[thick] (\i,\j) -- (\i+1,\j);
						}
						\foreach \i/\j in {0/0, 1/0, 0/-1, 1/-1, 0/-2, 1/-2}
						{
							\draw[thick] (\i,\j) -- (\i,\j-1);
						}
						\foreach \i/\j in {}
						{
							\draw[thick] (\i,\j) -- (\i-1,\j-1);
						}
					\end{tikzpicture}
				\end{array}$};
			\node(a4) at (3,0) {$\begin{array}{c}
					\begin{tikzpicture}[scale=0.3]
						\foreach \i/\j in {0/0, 0/-1, 1/0, 1/-1, 2/0}
						{
							\draw[thick] (\i,\j) -- (\i+1,\j);
						}
						\foreach \i/\j in {0/0, 1/0, 2/0, 0/-1}
						{
							\draw[thick] (\i,\j) -- (\i,\j-1);
						}
						\foreach \i/\j in {3/0, 1/-1}
						{
							\draw[thick] (\i,\j) -- (\i-1,\j-1);
						}
					\end{tikzpicture}
				\end{array}$};
			\node(a5) at (4.1,0) {$\begin{array}{c}
					\begin{tikzpicture}[scale=0.3]
						\foreach \i/\j in {0/0, 0/-1, 1/0, 0/-2}
						{
							\draw[thick] (\i,\j) -- (\i+1,\j);
						}
						\foreach \i/\j in {0/0, 1/0, 0/-1, 1/-1, 0/-2}
						{
							\draw[thick] (\i,\j) -- (\i,\j-1);
						}
						\foreach \i/\j in {2/0, 1/-2}
						{
							\draw[thick] (\i,\j) -- (\i-1,\j-1);
						}
					\end{tikzpicture}
				\end{array}$};
		\end{tikzpicture}
		\end{array} &
		\begin{array}{c}
			p_3, p_2 p_1, p_1^3, \theta_1 \theta_3, p_1 \theta_1 \theta_2
		\end{array} \\
		\hline
		\frac{7}{2} &  \begin{array}{c}\begin{tikzpicture}
			\node(a1) at (0,0) {$\begin{array}{c}
					\begin{tikzpicture}[scale=0.3]
						\foreach \i/\j in {0/0, 0/-1, 1/0, 1/-1, 2/0, 2/-1, 3/0}
						{
							\draw[thick] (\i,\j) -- (\i+1,\j);
						}
						\foreach \i/\j in {0/0, 1/0, 2/0, 3/0}
						{
							\draw[thick] (\i,\j) -- (\i,\j-1);
						}
						\foreach \i/\j in {4/0}
						{
							\draw[thick] (\i,\j) -- (\i-1,\j-1);
						}
					\end{tikzpicture}
				\end{array}$};
			\node(a2) at (1.4,0) {$\begin{array}{c}
					\begin{tikzpicture}[scale=0.3]
						\foreach \i/\j in {0/0, 0/-1, 1/0, 1/-1, 2/0, 2/-1}
						{
							\draw[thick] (\i,\j) -- (\i+1,\j);
						}
						\foreach \i/\j in {0/0, 1/0, 0/-1, 2/0, 3/0}
						{
							\draw[thick] (\i,\j) -- (\i,\j-1);
						}
						\foreach \i/\j in {1/-1}
						{
							\draw[thick] (\i,\j) -- (\i-1,\j-1);
						}
					\end{tikzpicture}
				\end{array}$};
			\node(a3) at (2.6,0) {$\begin{array}{c}
					\begin{tikzpicture}[scale=0.3]
						\foreach \i/\j in {0/0, 0/-1, 1/0, 1/-1, 2/0, 0/-2}
						{
							\draw[thick] (\i,\j) -- (\i+1,\j);
						}
						\foreach \i/\j in {0/0, 1/0, 0/-1, 2/0, 1/-1}
						{
							\draw[thick] (\i,\j) -- (\i,\j-1);
						}
						\foreach \i/\j in {3/0}
						{
							\draw[thick] (\i,\j) -- (\i-1,\j-1);
						}
					\end{tikzpicture}
				\end{array}$};
			\node(a4) at (3.65,0) {$\begin{array}{c}
					\begin{tikzpicture}[scale=0.3]
						\foreach \i/\j in {0/0, 0/-1, 0/-2, 1/0, 1/-1}
						{
							\draw[thick] (\i,\j) -- (\i+1,\j);
						}
						\foreach \i/\j in {0/0, 1/0, 0/-1, 1/-1, 2/0}
						{
							\draw[thick] (\i,\j) -- (\i,\j-1);
						}
						\foreach \i/\j in {2/-1}
						{
							\draw[thick] (\i,\j) -- (\i-1,\j-1);
						}
					\end{tikzpicture}
				\end{array}$};
			\node(a5) at (4.6,0) {$\begin{array}{c}
					\begin{tikzpicture}[scale=0.3]
						\foreach \i/\j in {0/0, 0/-1, 0/-2, 1/0, 1/-1}
						{
							\draw[thick] (\i,\j) -- (\i+1,\j);
						}
						\foreach \i/\j in {0/0, 1/0, 0/-1, 1/-1, 0/-2, 2/0}
						{
							\draw[thick] (\i,\j) -- (\i,\j-1);
						}
						\foreach \i/\j in {1/-2}
						{
							\draw[thick] (\i,\j) -- (\i-1,\j-1);
						}
					\end{tikzpicture}
				\end{array}$};
			\node(a6) at (5.6,0) {$\begin{array}{c}
					\begin{tikzpicture}[scale=0.3]
						\foreach \i/\j in {0/0, 0/-1, 0/-2, 1/0, 0/-3}
						{
							\draw[thick] (\i,\j) -- (\i+1,\j);
						}
						\foreach \i/\j in {0/0, 1/0, 0/-1, 1/-1, 0/-2, 1/-2}
						{
							\draw[thick] (\i,\j) -- (\i,\j-1);
						}
						\foreach \i/\j in {2/0}
						{
							\draw[thick] (\i,\j) -- (\i-1,\j-1);
						}
					\end{tikzpicture}
				\end{array}$};
			\node(a7) at (6.4,0) {$\begin{array}{c}
					\begin{tikzpicture}[scale=0.3]
						\foreach \i/\j in {0/0, 0/-1, 0/-2, 0/-3}
						{
							\draw[thick] (\i,\j) -- (\i+1,\j);
						}
						\foreach \i/\j in {0/0, 1/0, 0/-1, 1/-1, 0/-2, 1/-2, 0/-3}
						{
							\draw[thick] (\i,\j) -- (\i,\j-1);
						}
						\foreach \i/\j in {1/-3}
						{
							\draw[thick] (\i,\j) -- (\i-1,\j-1);
						}
					\end{tikzpicture}
				\end{array}$};
		\end{tikzpicture}
		\end{array} &
		\begin{array}{c}
			\theta_4, p_3 \theta_1, p_1 \theta_3, p_2 \theta_2, p_1 p_2 \theta_1, p_1^2 \theta_2, p_1^3 \theta_1
		\end{array} \\
		\hline
		4 &  \begin{array}{c}\begin{tikzpicture}
			\node(a1) at (0,0) {$\begin{array}{c}
					\begin{tikzpicture}[scale=0.3]
						\foreach \i/\j in {0/0, 0/-1, 1/0, 1/-1, 2/0, 2/-1, 3/0, 3/-1}
						{
							\draw[thick] (\i,\j) -- (\i+1,\j);
						}
						\foreach \i/\j in {0/0, 1/0, 2/0, 3/0, 4/0}
						{
							\draw[thick] (\i,\j) -- (\i,\j-1);
						}
						\foreach \i/\j in {}
						{
							\draw[thick] (\i,\j) -- (\i-1,\j-1);
						}
					\end{tikzpicture}
				\end{array}$};
			\node(a2) at (1.4,0) {$\begin{array}{c}
					\begin{tikzpicture}[scale=0.3]
						\foreach \i/\j in {0/0, 0/-1, 0/-2, 1/0, 1/-1, 2/0, 2/-1}
						{
							\draw[thick] (\i,\j) -- (\i+1,\j);
						}
						\foreach \i/\j in {0/0, 1/0, 0/-1, 1/-1, 2/0, 3/0}
						{
							\draw[thick] (\i,\j) -- (\i,\j-1);
						}
						\foreach \i/\j in {}
						{
							\draw[thick] (\i,\j) -- (\i-1,\j-1);
						}
					\end{tikzpicture}
				\end{array}$};
			\node(a3) at (2.7,0) {$\begin{array}{c}
					\begin{tikzpicture}[scale=0.3]
						\foreach \i/\j in {0/0, 0/-1, 0/-2, 1/0, 1/-1, 1/-2}
						{
							\draw[thick] (\i,\j) -- (\i+1,\j);
						}
						\foreach \i/\j in {0/0, 1/0, 0/-1, 1/-1, 2/0, 2/-1}
						{
							\draw[thick] (\i,\j) -- (\i,\j-1);
						}
						\foreach \i/\j in {}
						{
							\draw[thick] (\i,\j) -- (\i-1,\j-1);
						}
					\end{tikzpicture}
				\end{array}$};
			\node(a4) at (3.9,0) {$\begin{array}{c}
					\begin{tikzpicture}[scale=0.3]
						\foreach \i/\j in {0/0, 0/-1, 0/-2, 0/-3, 1/0, 1/-1}
						{
							\draw[thick] (\i,\j) -- (\i+1,\j);
						}
						\foreach \i/\j in {0/0, 1/0, 0/-1, 1/-1, 0/-2, 1/-2, 2/0}
						{
							\draw[thick] (\i,\j) -- (\i,\j-1);
						}
						\foreach \i/\j in {}
						{
							\draw[thick] (\i,\j) -- (\i-1,\j-1);
						}
					\end{tikzpicture}
				\end{array}$};
			\node(a5) at (4.7,0) {$\begin{array}{c}
					\begin{tikzpicture}[scale=0.3]
						\foreach \i/\j in {0/0, 0/-1, 0/-2, 0/-3, 0/-4}
						{
							\draw[thick] (\i,\j) -- (\i+1,\j);
						}
						\foreach \i/\j in {0/0, 1/0, 0/-1, 1/-1, 0/-2, 1/-2, 0/-3, 1/-3}
						{
							\draw[thick] (\i,\j) -- (\i,\j-1);
						}
						\foreach \i/\j in {}
						{
							\draw[thick] (\i,\j) -- (\i-1,\j-1);
						}
					\end{tikzpicture}
				\end{array}$};
			\node(a6) at (0,-1.5) {$\begin{array}{c}
					\begin{tikzpicture}[scale=0.3]
						\foreach \i/\j in {0/0, 0/-1, 1/0, 1/-1, 2/0, 2/-1, 3/0}
						{
							\draw[thick] (\i,\j) -- (\i+1,\j);
						}
						\foreach \i/\j in {0/0, 1/0, 2/0, 3/0, 0/-1}
						{
							\draw[thick] (\i,\j) -- (\i,\j-1);
						}
						\foreach \i/\j in {4/0, 1/-1}
						{
							\draw[thick] (\i,\j) -- (\i-1,\j-1);
						}
					\end{tikzpicture}
				\end{array}$};
			\node(a7) at (1.4,-1.5) {$\begin{array}{c}
					\begin{tikzpicture}[scale=0.3]
						\foreach \i/\j in {0/0, 0/-1, 0/-2, 1/0, 1/-1, 2/0}
						{
							\draw[thick] (\i,\j) -- (\i+1,\j);
						}
						\foreach \i/\j in {0/0, 1/0, 0/-1, 1/-1, 2/0}
						{
							\draw[thick] (\i,\j) -- (\i,\j-1);
						}
						\foreach \i/\j in {2/-1, 3/0}
						{
							\draw[thick] (\i,\j) -- (\i-1,\j-1);
						}
					\end{tikzpicture}
				\end{array}$};
			\node(a8) at (2.7,-1.5) {$\begin{array}{c}
					\begin{tikzpicture}[scale=0.3]
						\foreach \i/\j in {0/0, 0/-1, 1/0, 1/-1, 2/0, 0/-2}
						{
							\draw[thick] (\i,\j) -- (\i+1,\j);
						}
						\foreach \i/\j in {0/0, 1/0, 0/-1, 2/0, 1/-1, 0/-2}
						{
							\draw[thick] (\i,\j) -- (\i,\j-1);
						}
						\foreach \i/\j in {3/0, 1/-2}
						{
							\draw[thick] (\i,\j) -- (\i-1,\j-1);
						}
					\end{tikzpicture}
				\end{array}$};
			\node(a9) at (3.8,-1.5) {$\begin{array}{c}
					\begin{tikzpicture}[scale=0.3]
						\foreach \i/\j in {0/0, 0/-1, 0/-2, 1/0, 1/-1}
						{
							\draw[thick] (\i,\j) -- (\i+1,\j);
						}
						\foreach \i/\j in {0/0, 1/0, 0/-1, 1/-1, 2/0, 0/-2}
						{
							\draw[thick] (\i,\j) -- (\i,\j-1);
						}
						\foreach \i/\j in {2/-1, 1/-2}
						{
							\draw[thick] (\i,\j) -- (\i-1,\j-1);
						}
					\end{tikzpicture}
				\end{array}$};
			\node(a10) at (4.9,-1.5) {$\begin{array}{c}
					\begin{tikzpicture}[scale=0.3]
						\foreach \i/\j in {0/0, 0/-1, 0/-2, 0/-3, 1/0}
						{
							\draw[thick] (\i,\j) -- (\i+1,\j);
						}
						\foreach \i/\j in {0/0, 1/0, 0/-1, 1/-1, 0/-2, 1/-2, 0/-3}
						{
							\draw[thick] (\i,\j) -- (\i,\j-1);
						}
						\foreach \i/\j in {1/-3, 2/0}
						{
							\draw[thick] (\i,\j) -- (\i-1,\j-1);
						}
					\end{tikzpicture}
				\end{array}$};
		\end{tikzpicture}\end{array} &
		\begin{array}{c}
			p_4, p_3 p_1, p_2^2, p_2 p_1^2, p_1^4, \\
			\\
			\theta_1 \theta_4, \theta_2 \theta_3, p_1 \theta_1 \theta_3, p_2 \theta_1 \theta_2, p_1^2 \theta_1 \theta_2
		\end{array} \\
		\hline
	\end{array}$
	\caption{Super-Young diagrams are in one-to-one correspondence to the homogeneous polynomials in double set of variables $p_a$ and $\theta_a$.}
	\label{Fermion table}
\end{table}

This paper is organized as follows. In Section \ref{sec:Schur} we provide basic facts about usual Schur polynomials and discuss a family of commuting operators that can independently define Schur polynomials as common set of eigenfunctions. In Section \ref{sec:Macdonald} we discuss various properties of canonical Macdonald polynomials. In Section \ref{sec:super-Schur} we define analog of Schur polynomials for super-partitions via commuting set of operators from affine super-Yangian $Y(\hat{\mathfrak{gl}}_{1|1})$. In the main Section \ref{sec:super-Macdonald} we define super-Macdonald polynomials and check various properties of these new polynomials in Section \ref{sec:Properties}. We summarize the ideas in a concise conclusion in Section \ref{sec:Conclusion}.

\section{Schur polynomials} \label{sec:Schur}

Schur polynomials \cite{Macdonald} arise in various areas of modern mathematical and theoretical physics, however one of the main properties is that they are $GL(N)$ characters. Tensor product of two irreducible $GL(N)$ representation labeled by Young diagrams $\lambda$ and $\lambda^{\prime}$ breaks into direct sum of other irreducible representations $\mu$:
\begin{equation}
	\lambda \otimes \lambda^{\prime} = \oplus_{\mu} C_{\lambda, \lambda^{\prime}}^{\mu} \, \mu
\end{equation}
where integer Littlewood-Richardson coefficients $C_{\lambda, \lambda^{\prime}}^{\mu}$ counts how many isomorphic representations are included in the sum. Schur polynomials $S_{\lambda}$ satisfy exactly the same relations:
\begin{equation}
	S_{\lambda} \cdot S_{\lambda^{\prime}} = \sum_{\mu} C_{\lambda, \lambda^{\prime}}^{\mu} \,  S_{\mu}
\end{equation}
We use Schur polynomials as functions of $p_a$ variable where $a \in \mathbb{N}$ is a natural number. To proceed to realization of Schur polynomials as symmetric functions one can perform Miwa change of variables $p_k = \sum_{i=1}^{N} (x_i)^k$.

There are many ways to compute Schur polynomials and we list here two of them. The first way is the simplest and it includes two steps. On the first step we compute Schur polynomials corresponding to one-row diagrams $[n]$ from the following generating function:
\begin{equation}
	\exp\left( \sum_{n=1} \frac{p_n}{n} z^n \right) = \sum_{n = 0} z^n \, S_{[n]}  = 1 + z \, p_1 + z^2 \left( \frac{p_1^2}{2} + \frac{p_2}{2} \right) + \ldots
\end{equation}
Then Schur polynomial for arbitrary Young diagram $\lambda$ can be computed via famous Jacobi-Trudi formula:
\begin{equation}
	S_{\lambda} = \det_{1 \leqslant i, j \leqslant l(\lambda)} \left( S_{[\lambda_i + j - i]}\right)
\end{equation}

Explicit examples of Schur polynomials from small levels read:
\begin{align}
	\begin{aligned}
		S_{\varnothing} = 1 \hspace{15mm} S_{\, \begin{tikzpicture}[scale=0.15]
				\foreach \i/\j in {0/0, 0/-1}
				{
					\draw[thick] (\i,\j) -- (\i+1,\j);
				}
				\foreach \i/\j in {0/0, 1/0}
				{
					\draw[thick] (\i,\j) -- (\i,\j-1);
				}
				\foreach \i/\j in {}
				{
					\draw[thick] (\i,\j) -- (\i-1,\j-1);
				}
		\end{tikzpicture}} = p_1 \hspace{15mm} 
		S_{\, \begin{tikzpicture}[scale=0.15]
				\foreach \i/\j in {0/0, 0/-1, 1/0, 1/-1}
				{
					\draw[thick] (\i,\j) -- (\i+1,\j);
				}
				\foreach \i/\j in {0/0, 1/0, 2/0}
				{
					\draw[thick] (\i,\j) -- (\i,\j-1);
				}
				\foreach \i/\j in {}
				{
					\draw[thick] (\i,\j) -- (\i-1,\j-1);
				}
		\end{tikzpicture}} = \frac{p_1^2}{2} + \frac{p_2}{2} \hspace{15mm}
		S_{\, \begin{tikzpicture}[scale=0.15]
				\foreach \i/\j in {0/0, 0/-1, 0/-2}
				{
					\draw[thick] (\i,\j) -- (\i+1,\j);
				}
				\foreach \i/\j in {0/0, 1/0, 0/-1, 1/-1}
				{
					\draw[thick] (\i,\j) -- (\i,\j-1);
				}
				\foreach \i/\j in {}
				{
					\draw[thick] (\i,\j) -- (\i-1,\j-1);
				}
		\end{tikzpicture}} = \frac{p_1^2}{2} - \frac{p_2}{2} \\
		S_{\, \begin{tikzpicture}[scale=0.15]
				\foreach \i/\j in {0/0, 0/-1, 1/0, 1/-1, 2/0, 2/-1}
				{
					\draw[thick] (\i,\j) -- (\i+1,\j);
				}
				\foreach \i/\j in {0/0, 1/0, 2/0, 3/0}
				{
					\draw[thick] (\i,\j) -- (\i,\j-1);
				}
				\foreach \i/\j in {}
				{
					\draw[thick] (\i,\j) -- (\i-1,\j-1);
				}
		\end{tikzpicture}} = \frac{p_1^3}{6} + \frac{p_1 p_2}{2} + \frac{p_3}{3} 
		\hspace{15mm}
			S_{\, \begin{tikzpicture}[scale=0.15]
			\foreach \i/\j in {0/0, 0/-1, 1/0, 1/-1, 0/-2}
			{
				\draw[thick] (\i,\j) -- (\i+1,\j);
			}
			\foreach \i/\j in {0/0, 1/0, 2/0, 0/-1, 1/-1}
			{
				\draw[thick] (\i,\j) -- (\i,\j-1);
			}
			\foreach \i/\j in {}
			{
				\draw[thick] (\i,\j) -- (\i-1,\j-1);
			}
		\end{tikzpicture}} = \frac{p_1^3}{3} -\frac{p_3}{3}
		\hspace{15mm}
		S_{\, \begin{tikzpicture}[scale=0.15]
		\foreach \i/\j in {0/0, 0/-1, 0/-2, 0/-3}
		{
			\draw[thick] (\i,\j) -- (\i+1,\j);
		}
		\foreach \i/\j in {0/0, 1/0, 0/-1, 1/-1, 0/-2, 1/-2}
		{
			\draw[thick] (\i,\j) -- (\i,\j-1);
		}
		\foreach \i/\j in {}
		{
			\draw[thick] (\i,\j) -- (\i-1,\j-1);
		}
		\end{tikzpicture}} = \frac{p_1^3}{6} - \frac{p_1 p_2}{2} + \frac{p_3}{3}
	\end{aligned}
\end{align}

Schur polynomials can be defined as common eigenfunctions of commuting set of operators $\hat{\psi}_n$, where $n = 0,1,2,3,\ldots$:
\begin{align}
	\hat{\psi}_n \, S_{\lambda}(p) = \mathcal{E}_{n,\lambda} \, S_{\lambda}(p)
\end{align}
These operators are Cartan operators of infinite dimensional affine Yangian $Y(\hat{\mathfrak{gl}_1})$ in Fock representation \cite{Prochazka:2015deb, Tsymbaliuk2017, Morozov:2023vra}. These operators are also lie in bigger family of cut-and-join operators \cite{Mironov:2009cj, ivanov2003, Mironov:2010yg, Mironov:2019mah} that play an important role in Schur-Weyl duality. The set of operators $\hat{\psi}_n$ is constructed in four steps:
\begin{enumerate}
	\item On the first step we define three auxiliary operators:
	\begin{align}
		\hat{e}_0 := p_1, \hspace{10mm} \hat{f}_0 := \frac{\partial}{\partial p_1} \hspace{10mm} \hat{W} := \frac{1}{2}\sum_{a,b= 1}^{\infty}ab \,p_{a+b}\frac{\partial^2}{\partial p_a \partial p_b}+(a+b)p_ap_b\frac{\partial}{\partial p_{a+b}}
	\end{align}
	\item On the second step we compute higher auxiliary operators recursively from the initial ones:
	\begin{align}
		\hat{e}_{n+1} := \Big[ \hat{W}, \hat{e}_{n}\Big]
	\end{align}
	\item On the third step we compute commutators of $\hat{e}_n, \hat{f}_0$ that give the resulting set of commuting operators:
	\begin{align}
		\label{psi n Yangian}
		\hat{\psi}_{n} := \Big[ \hat{e}_{n}, \hat{f}_{0} \Big] \hspace{20mm} \Big[ \hat{\psi}_n, \hat{\psi}_m \Big] = 0 
	\end{align}
	In particular, $\hat{W}$ operator is a member of this set:
	\begin{equation}
		\hat{W} = - \frac{1}{6} \hat{\psi}_3
	\end{equation}
	\item On the final step we compute eigenvalues $\mathcal{E}_{n,\lambda}$ from the representation theory of affine Yangian \cite{Prochazka:2015deb}. These eigenvalues are collected in the generating functions:
	\begin{equation}
		\mathcal{E}_{\lambda}(z) = 1 - (h_1 + h_2) \sum_{n=0} \frac{\mathcal{E}_{n,\lambda}(h_1, h_2)}{z^{n+1}}, \hspace{10mm} \mathcal{E}_{n,\lambda} = \mathcal{E}_{n,\lambda}(h_1 = 1, h_2 = -1)
	\end{equation}
	Note that we firstly compute the expansion coefficients and then evaluate it at special point $h_1 = 1, h_2 = -1$. We do it because of factor $(h_1 + h_2)$ that vanishes at this special point. The generating function itself is computed as follows:
	\begin{equation}
		\mathcal{E}_{\lambda}(z) = \frac{h_1 + h_2 + z}{z} \cdot \prod_{\begin{tikzpicture}[scale=0.15]
				\foreach \i/\j in {0/0, 0/-1}
				{
					\draw[thick] (\i,\j) -- (\i+1,\j);
				}
				\foreach \i/\j in {0/0, 1/0}
				{
					\draw[thick] (\i,\j) -- (\i,\j-1);
				}
				\foreach \i/\j in {}
				{
					\draw[thick] (\i,\j) -- (\i-1,\j-1);
				}
			\end{tikzpicture} \, \in \lambda} \varphi\left( z - \Omega_{\begin{tikzpicture}[scale=0.15]
			\foreach \i/\j in {0/0, 0/-1}
			{
				\draw[thick] (\i,\j) -- (\i+1,\j);
			}
			\foreach \i/\j in {0/0, 1/0}
			{
				\draw[thick] (\i,\j) -- (\i,\j-1);
			}
			\foreach \i/\j in {}
			{
				\draw[thick] (\i,\j) -- (\i-1,\j-1);
			}
		\end{tikzpicture}} \right)
	\end{equation}
	In this formula we introduced content $\Omega = h_1 \cdot j + h_2 \cdot i$, where $(i,j)$ is vertical and horizontal coordinates of a box, starting box has coordinates $(0,0)$. The function $\varphi$ has the following form:
	\begin{equation}
		\varphi(z) := \frac{(z + h_1)(z + h_2)(z - h_1 - h_2)}{(z - h_1)(z - h_2)(z + h_1 + h_2)}
	\end{equation}
	For illustrative purposes we provide an example:
	\begin{equation}
		\begin{aligned}
			\mathcal{E}_{\, \begin{tikzpicture}[scale=0.15]
					\foreach \i/\j in {0/0, 0/-1, 0/-2, 1/0, 1/-1, 2/0, 2/-1}
					{
						\draw[thick] (\i,\j) -- (\i+1,\j);
					}
					\foreach \i/\j in {0/0, 1/0, 0/-1, 1/-1, 2/0, 3/0}
					{
						\draw[thick] (\i,\j) -- (\i,\j-1);
					}
					\foreach \i/\j in {}
					{
						\draw[thick] (\i,\j) -- (\i-1,\j-1);
					}
			\end{tikzpicture}} (z) &= \frac{h_1 + h_2 + z}{z} \varphi(z) \varphi(z - h_1) \varphi(z - 2h_1) \varphi(z - h_2) = \\
		  &= 1 - (h_1 + h_2) \left( - \frac{1}{z} + \frac{8 \left(h_1 h_2\right)}{z^3} + \frac{2 \left(h_1 h_2 \left(7 h_1+13 h_2\right)\right)}{z^4} + \ldots \right)
		\end{aligned}
	\end{equation}
	Evaluating the expansion coefficients at the special point $h_1 = 1, h_2 = -1$ we obtain the following eigenvalues:
	\begin{equation}
		\mathcal{E}_{0, \, \begin{tikzpicture}[scale=0.15]
				\foreach \i/\j in {0/0, 0/-1, 0/-2, 1/0, 1/-1, 2/0, 2/-1}
				{
					\draw[thick] (\i,\j) -- (\i+1,\j);
				}
				\foreach \i/\j in {0/0, 1/0, 0/-1, 1/-1, 2/0, 3/0}
				{
					\draw[thick] (\i,\j) -- (\i,\j-1);
				}
				\foreach \i/\j in {}
				{
					\draw[thick] (\i,\j) -- (\i-1,\j-1);
				}
		\end{tikzpicture}} = -1, \hspace{5mm}
		\mathcal{E}_{1, \, \begin{tikzpicture}[scale=0.15]
				\foreach \i/\j in {0/0, 0/-1, 0/-2, 1/0, 1/-1, 2/0, 2/-1}
				{
					\draw[thick] (\i,\j) -- (\i+1,\j);
				}
				\foreach \i/\j in {0/0, 1/0, 0/-1, 1/-1, 2/0, 3/0}
				{
					\draw[thick] (\i,\j) -- (\i,\j-1);
				}
				\foreach \i/\j in {}
				{
					\draw[thick] (\i,\j) -- (\i-1,\j-1);
				}
		\end{tikzpicture}} = 0, \hspace{5mm}
		\mathcal{E}_{2, \, \begin{tikzpicture}[scale=0.15]
				\foreach \i/\j in {0/0, 0/-1, 0/-2, 1/0, 1/-1, 2/0, 2/-1}
				{
					\draw[thick] (\i,\j) -- (\i+1,\j);
				}
				\foreach \i/\j in {0/0, 1/0, 0/-1, 1/-1, 2/0, 3/0}
				{
					\draw[thick] (\i,\j) -- (\i,\j-1);
				}
				\foreach \i/\j in {}
				{
					\draw[thick] (\i,\j) -- (\i-1,\j-1);
				}
		\end{tikzpicture}} = -8, \hspace{5mm}
		\mathcal{E}_{3, \, \begin{tikzpicture}[scale=0.15]
				\foreach \i/\j in {0/0, 0/-1, 0/-2, 1/0, 1/-1, 2/0, 2/-1}
				{
					\draw[thick] (\i,\j) -- (\i+1,\j);
				}
				\foreach \i/\j in {0/0, 1/0, 0/-1, 1/-1, 2/0, 3/0}
				{
					\draw[thick] (\i,\j) -- (\i,\j-1);
				}
				\foreach \i/\j in {}
				{
					\draw[thick] (\i,\j) -- (\i-1,\j-1);
				}
		\end{tikzpicture}} = 12
	\end{equation}
\end{enumerate}

\bigskip

Another important property of Schur polynomials is that they form orthonormal set:
\begin{equation}
	\bra{S_{\lambda}} \ket{S_{\mu} } = \delta_{\lambda, \mu}
\end{equation}
with respect to the following scalar product defined for basis monomials:
\begin{equation}
	\bra{p_{\Delta}} \ket{p_{\Delta^{\prime}}} = \delta_{\Delta, \Delta^{\prime}} \cdot z_{\Delta}
\end{equation}
In the above formula we use the following notations. $\Delta = \left[ \Delta_1, \Delta_2, \ldots, \Delta_{l(\Delta)}\right]$ is a Young diagram. Basis monomials are labeled by the Young diagrams $p_{\Delta} := \prod_{k=1}^{l(\Delta)} p_{\Delta_k}$. Combinatorial factors $z_{\Delta}$ are best understood in dual presentation $\Delta = \left\{ 1^{m_1}, 2^{m_2}, 3^{m_3}, \ldots \right\}$ of Young diagrams.
This notation means that $\Delta$ contains $m_1$ rows of unit length, $m_2$ rows of length two and so on. Then for the $z_{\Delta}$ the following formula holds $ z_{\Delta} = \prod_{k=1} \, m_k! \cdot k^{m_k}$.
Equivalently, the orthogonality relations can be represented via Cauchy identity \cite{Macdonald, MorozovCauchy} in an elegant way:
\begin{equation}
	\sum_{\lambda} S_{\lambda} (p) \cdot S_{\lambda}(\bar{p}) = \exp\left( \sum_{n=1} \frac{p_n \, \bar{p}_n}{n} \right)
\end{equation}
In the above formula the sum in the l.h.s. runs over all Young diagrams and $\bar{p}_k$ are the second auxiliary copy of $p_k$ variables. Cauchy identity has another useful form:
\begin{equation}
	\sum_{\lambda} \, (-)^{|\lambda|} \, S_{\lambda} (p) \cdot S_{\lambda^{T}}(-\bar{p}) = \exp\left( \sum_{n=1} \frac{p_n \, \bar{p}_n}{n} \right)
\end{equation}
that dictates the transposition rule for Schur polynomials:
\begin{equation}
	S_{\lambda^{T}}(p_k) = (-)^{|\lambda|} \, S_{\lambda} (-p_k)
\end{equation}
Here we use the notation $\lambda^T$ for transposed diagram $\lambda$ and $|\lambda|$ for the total number of boxes in the corresponding diagram.
\section{From Schur to Macdonald} \label{sec:Macdonald}

Macdonald polynomials \cite{Macdonald} $M^{q,t}_{\lambda} (p)$ is a very special family of $(q,t)$-deformed polynomials and can be defined in different ways due to its various properties. One of the simplest ways is to define Macdonald polynomials as eigenfunctions of the first \cite{RUIJSENAARS1986, ruijsenaars1987complete} Hamiltonian (for review see \cite{Mironov:2019uoy}):

\begin{equation}
	\label{Macdonald H}
	\hat{H} = \oint \frac{dz}{z} \exp \left( \sum_{k = 1} \frac{(1-t^{-2k})}{k} \, p_k \, z^k \right) \, \exp \left(\sum_{k = 1} \frac{(q^{2k} - 1)}{z^k} \, \frac{\partial}{\partial p_k} \right)
\end{equation}

\begin{align}
	\hat{H} \, M^{q,t}_{\lambda} = \mathcal{E}^{\, q,t}_{\lambda} \, M^{q,t}_{\lambda}
\end{align}
For higher Hamiltonians see  \cite{Mironov:2019uoy} and \cite{Mironov:2024sbc}. The formula for eigenvalue $\mathcal{E}^{\, q,t}_{\lambda}$ has the form of the sum over all boxes in the Young diagram:
\begin{equation}
	\mathcal{E}^{\, q,t}_{\lambda} = 1 + (q^2 + 1)(1 - t^{-2}) \cdot \sum_{\Box \in \lambda} q^{2 j_{\Box}} t^{-2 i_{\Box}}
\end{equation}
where the coordinate of the initial box in the Young diagram has coordinates $(i_{\Box}, j_{\Box}) = (0,0)$ and $i_{\Box}$, $j_{\Box}$ are vertical and horizontal coordinates correspondingly. Remarkably, there are no degeneracy in eigenvalues, therefore the only one operator is enough to define the whole set of Macdonald polynomials. We list Macdonald polynomials from small levels:

\begin{align}
	\begin{aligned}
		M^{q,t}_{\varnothing} &= 1 \\
		M^{q,t}_{\, \begin{tikzpicture}[scale=0.15]
				\foreach \i/\j in {0/0, 0/-1}
				{
					\draw[thick] (\i,\j) -- (\i+1,\j);
				}
				\foreach \i/\j in {0/0, 1/0}
				{
					\draw[thick] (\i,\j) -- (\i,\j-1);
				}
				\foreach \i/\j in {}
				{
					\draw[thick] (\i,\j) -- (\i-1,\j-1);
				}
		\end{tikzpicture}} &= p_1 \\
		M^{q,t}_{\, \begin{tikzpicture}[scale=0.15]
				\foreach \i/\j in {0/0, 0/-1, 1/0, 1/-1}
				{
					\draw[thick] (\i,\j) -- (\i+1,\j);
				}
				\foreach \i/\j in {0/0, 1/0, 2/0}
				{
					\draw[thick] (\i,\j) -- (\i,\j-1);
				}
				\foreach \i/\j in {}
				{
					\draw[thick] (\i,\j) -- (\i-1,\j-1);
				}
		\end{tikzpicture}} &= \frac{\left(q^2-1\right) \left(t^2+1\right) p_2}{ 2(q^2 t^2-1)} + \frac{\left(q^2+1\right) \left(t^2-1\right) p_1^2}{2 (q^2 t^2-1)} \\
		M^{q,t}_{\, \begin{tikzpicture}[scale=0.15]
				\foreach \i/\j in {0/0, 0/-1, 0/-2}
				{
					\draw[thick] (\i,\j) -- (\i+1,\j);
				}
				\foreach \i/\j in {0/0, 1/0, 0/-1, 1/-1}
				{
					\draw[thick] (\i,\j) -- (\i,\j-1);
				}
				\foreach \i/\j in {}
				{
					\draw[thick] (\i,\j) -- (\i-1,\j-1);
				}
		\end{tikzpicture}} &= \frac{p_1^2}{2} - \frac{p_2}{2} \\
		M^{q,t}_{\, \begin{tikzpicture}[scale=0.15]
				\foreach \i/\j in {0/0, 0/-1, 1/0, 1/-1, 2/0, 2/-1}
				{
					\draw[thick] (\i,\j) -- (\i+1,\j);
				}
				\foreach \i/\j in {0/0, 1/0, 2/0, 3/0}
				{
					\draw[thick] (\i,\j) -- (\i,\j-1);
				}
				\foreach \i/\j in {}
				{
					\draw[thick] (\i,\j) -- (\i-1,\j-1);
				}
		\end{tikzpicture}} &= \frac{\left(q^2-1\right)^2 \left(q^2+1\right) \left(t^4+t^2+1\right)p_3}{3 \left(q^2 t^2-1\right) \left(q^4 t^2-1\right)}  + \frac{\left(q^6-1\right) \left(t^4-1\right)p_2 p_1}{2 \left(q^2 t^2-1\right) \left(q^4 t^2-1\right)}  + \frac{\left(q^6+2 q^4+2 q^2+1\right) \left(t^2-1\right)^2 p_1^3}{6 \left(q^2 t^2-1\right) \left(q^4 t^2-1\right)} \\
		M^{q,t}_{\, \begin{tikzpicture}[scale=0.15]
				\foreach \i/\j in {0/0, 0/-1, 1/0, 1/-1, 0/-2}
				{
					\draw[thick] (\i,\j) -- (\i+1,\j);
				}
				\foreach \i/\j in {0/0, 1/0, 2/0, 0/-1, 1/-1}
				{
					\draw[thick] (\i,\j) -- (\i,\j-1);
				}
				\foreach \i/\j in {}
				{
					\draw[thick] (\i,\j) -- (\i-1,\j-1);
				}
		\end{tikzpicture}} &= \frac{2 \left(q^2-1\right) \left(t^4+t^2+1\right)p_3}{6(1- q^2 t^4)} + \frac{\left(t^2+1\right) \left(q^2-t^2\right) p_2 p_1}{2 (q^2 t^4-1)} + \frac{\left(t^2-1\right) \left(2 q^2 t^2+ q^2 +t^2+2\right)p_1^3}{6( q^2 t^4-1)} \\
		M^{q,t}_{\, \begin{tikzpicture}[scale=0.15]
				\foreach \i/\j in {0/0, 0/-1, 0/-2, 0/-3}
				{
					\draw[thick] (\i,\j) -- (\i+1,\j);
				}
				\foreach \i/\j in {0/0, 1/0, 0/-1, 1/-1, 0/-2, 1/-2}
				{
					\draw[thick] (\i,\j) -- (\i,\j-1);
				}
				\foreach \i/\j in {}
				{
					\draw[thick] (\i,\j) -- (\i-1,\j-1);
				}
		\end{tikzpicture}} &= \frac{p_1^3}{6} - \frac{p_1 p_2}{2} + \frac{p_3}{3}\\
	\end{aligned}
\end{align}

Macdonald polynomials give Jack polynomials in the limit $t = e^{\beta \hbar}$, $q = e^{\hbar}$, $\hbar \to 0$, $r$-Uglov polynomials \cite{Uglov:1997ia, Galakhov:2024mbz, Mishnyakov:2024cgl} in the limit $t = e^{\beta \hbar + \pi i / r}$, $q = e^{\hbar  + \pi i / r}$, $\hbar \to 0$, Schur polynomials in the limit $q = t$.

In the final part of this section about Macdonald polynomials we list their distinguished properties: \\
	$\bullet$ \underline{Orthogonality relations}. Macdonald polynomials are orthogonal with respect to the following scalar product:
	\begin{equation}
		\bra{M^{q,t}_{\lambda}} \ket{M^{q,t}_{\mu} } = || M^{q,t}_{\lambda} ||^2 \cdot \delta_{\lambda, \mu}
	\end{equation}
	provided that the scalar product of basis monomials reads:
	\begin{equation}
		\label{q,t measure Macdonald}
		\bra{p_{\Delta}} \ket{p_{\Delta^{\prime}}} = \delta_{\Delta, \Delta^{\prime}} \cdot z_{\Delta} \cdot \prod_{k=1}^{l(\Delta)} \frac{q^{2 \Delta_k} - 1}{t^{2 \Delta_k} - 1}
	\end{equation}
	
	The orthogonality relations can be equivalently represented by the generalized Cauchy identity:
	\begin{tcolorbox}
		\begin{equation}
			\label{Cauchy Macdonald}
			\sum_{\lambda} \ \frac{M^{q,t}_{\lambda}(p) \cdot M^{q,t}_{\lambda}(\bar{p})}{|| M^{q,t}_{\lambda}||^2} = \exp\left( \sum_{n=1} \frac{p_n \, \bar{p}_n}{n} \, \frac{t^{2n}-1}{q^{2n}-1} \right)
		\end{equation}
	\end{tcolorbox}
	$\bullet$ \underline{Triangular decompositions}. The set Young diagrams has natural ordering $\succ_{r}$ by lexicographical rule with respect of rows (hence letter $r$ the label $\succ_{r}$): for any pair of diagrams $\lambda \not= \mu$ of equal number of boxes $|\lambda| = |\mu|$, $\lambda \succ_{r} \mu$ if one of the following statements is fulfilled:
	\begin{enumerate}
		\item $\lambda_1 > \mu_1$
		\item $\lambda_1 = \mu_1$, $\lambda_2 > \mu_2$
		\item $\lambda_1 = \mu_1, \lambda_2 = \mu_2$, $\lambda_3 > \mu_3$
		\item \ldots
	\end{enumerate}
	Completely analogically the other label $\prec_{r}$ is defined. Macdonald polynomials are expanded in the basis of Schur polynomials by the following triangular formula that extensively uses the above ordering:
	\begin{tcolorbox}
		\begin{equation}
			\label{row triangularity Macdonald}
			M_{\lambda}^{q,t} = S_{\lambda} + \sum_{\mu \, \prec_{r} \, \lambda} K_{\lambda, \mu}^{q,t} \, S_{\mu}
		\end{equation}
	\end{tcolorbox}
	It is implied that the sum goes over diagrams $\mu$ of the same size as $\lambda$, i.e. $|\mu| = |\lambda|$. The triangular property strongly fixes the form of Macdonald polynomials. For example, for Young diagrams containing only one column it fixes the corresponding Macdonald polynomials completely since the one column diagrams is the lowest diagram according to the lexicographical order and the sum over $\mu$ is empty:
	\begin{equation}
		M^{q,t}_{[1^n]} = S_{[1^n]}
	\end{equation}
	Starting from Schur polynomials and taking \boxed{ \textbf{ orthogonality \eqref{Cauchy Macdonald} and triangularity \eqref{row triangularity Macdonald} }} as defining properties \cite{Mironov:2020aaa} one can fully determine the form of Macdonald polynomials. Our plan is to define the super-Macdonald polynomials according to this fact. We discuss the plan in detail in Section \ref{sec:super-Macdonald}.
	
	Remarkably, Macdonald polynomials respect also \textit{another one} triangular decomposition. The second decomposition is based on lexicographical order with respect to columns, in contrast with the first one that involves ordering with respect to rows. This ordering is defined in the following way. For any pair of diagrams $\lambda \not= \mu$ of equal number of boxes $|\lambda| = |\mu|$, $\lambda \succ_{c} \mu$ if one of the following statements is fulfilled:
	\begin{enumerate}
		\item $\lambda^T_1 < \mu^T_1$
		\item $\lambda^T_1 = \mu^T_1$, $\lambda^T_2 < \mu^T_2$
		\item $\lambda^T_1 = \mu^T_1, \lambda^T_2 = \mu^T_2$, $\lambda^T_3 < \mu^T_3$
		\item \ldots
	\end{enumerate}
	Note two main differences from the first ordering. Firstly, we compare diagrams by column and use notation $\lambda^T$ that means transposed diagram. In other words, $\lambda^{T}_k$ is the length of the $k$-th column of the diagram $\lambda$. Secondly, note that we define $\lambda \succ_{c} \mu$ if $\lambda^T_1 < \mu^T_1$, i.e. the order is inverted comparing to the row-ordering where $\lambda \succ_{r} \mu$ if $\lambda_1 > \mu_1$. These naively confusing rules turns out to be convenient: the row and column ordering almost always coincide that makes it easy to compare them. The orderings start to differ from the $6$-th level. We provide examples on $5$-th and $6$-th levels for illustrative purposes.
	
	$5$-th level:
	\begin{equation}
		{\begin{array}{c}
				\begin{tikzpicture}[scale=0.3]
					\foreach \i/\j in {0/0, 1/0, 2/0, 3/0, 4/0, 0/-1, 1/-1, 2/-1, 3/-1, 4/-1}
					{
						\draw[thick] (\i,\j) -- (\i+1,\j);
					}
					\foreach \i/\j in {0/0, 1/0, 2/0, 3/0, 4/0, 5/0}
					{
						\draw[thick] (\i,\j) -- (\i,\j-1);
					}
					\foreach \i/\j in {}
					{
						\draw[thick] (\i,\j) -- (\i-1,\j-1);
					}
				\end{tikzpicture}
		\end{array}} \succ_{r,c}
		{\begin{array}{c}
				\begin{tikzpicture}[scale=0.3]
					\foreach \i/\j in {0/0, 1/0, 2/0, 3/0, 0/-1, 1/-1, 2/-1, 3/-1, 0/-2}
					{
						\draw[thick] (\i,\j) -- (\i+1,\j);
					}
					\foreach \i/\j in {0/0, 1/0, 2/0, 3/0, 4/0, 0/-1, 1/-1}
					{
						\draw[thick] (\i,\j) -- (\i,\j-1);
					}
					\foreach \i/\j in {}
					{
						\draw[thick] (\i,\j) -- (\i-1,\j-1);
					}
				\end{tikzpicture}
		\end{array}} \succ_{r,c}
		{\begin{array}{c}
				\begin{tikzpicture}[scale=0.3]
					\foreach \i/\j in {0/0, 1/0, 2/0, 0/-1, 1/-1, 2/-1, 0/-2, 1/-2}
					{
						\draw[thick] (\i,\j) -- (\i+1,\j);
					}
					\foreach \i/\j in {0/0, 1/0, 2/0, 3/0, 0/-1, 1/-1, 2/-1}
					{
						\draw[thick] (\i,\j) -- (\i,\j-1);
					}
					\foreach \i/\j in {}
					{
						\draw[thick] (\i,\j) -- (\i-1,\j-1);
					}
				\end{tikzpicture}
		\end{array}} \succ_{r,c}
		{\begin{array}{c}
				\begin{tikzpicture}[scale=0.3]
					\foreach \i/\j in {0/0, 1/0, 2/0, 0/-1, 1/-1, 2/-1, 0/-2, 0/-3}
					{
						\draw[thick] (\i,\j) -- (\i+1,\j);
					}
					\foreach \i/\j in {0/0, 1/0, 2/0, 3/0, 0/-1, 1/-1, 0/-2, 1/-2}
					{
						\draw[thick] (\i,\j) -- (\i,\j-1);
					}
					\foreach \i/\j in {}
					{
						\draw[thick] (\i,\j) -- (\i-1,\j-1);
					}
				\end{tikzpicture}
		\end{array}} \succ_{r,c}
		{\begin{array}{c}
				\begin{tikzpicture}[scale=0.3]
					\foreach \i/\j in {0/0, 1/0, 0/-1, 1/-1, 0/-2, 0/-3, 1/-2}
					{
						\draw[thick] (\i,\j) -- (\i+1,\j);
					}
					\foreach \i/\j in {0/0, 1/0, 2/0, 0/-1, 1/-1, 0/-2, 1/-2, 2/-1}
					{
						\draw[thick] (\i,\j) -- (\i,\j-1);
					}
					\foreach \i/\j in {}
					{
						\draw[thick] (\i,\j) -- (\i-1,\j-1);
					}
				\end{tikzpicture}
		\end{array}} \succ_{r,c}
		{\begin{array}{c}
				\begin{tikzpicture}[scale=0.3]
					\foreach \i/\j in {0/0, 1/0, 0/-1, 1/-1, 0/-2, 0/-3, 0/-4}
					{
						\draw[thick] (\i,\j) -- (\i+1,\j);
					}
					\foreach \i/\j in {0/0, 1/0, 2/0, 0/-1, 1/-1, 0/-2, 1/-2, 0/-3, 1/-3}
					{
						\draw[thick] (\i,\j) -- (\i,\j-1);
					}
					\foreach \i/\j in {}
					{
						\draw[thick] (\i,\j) -- (\i-1,\j-1);
					}
				\end{tikzpicture}
		\end{array}} \succ_{r,c}
		{\begin{array}{c}
				\begin{tikzpicture}[scale=0.3]
					\foreach \i/\j in {0/0, 0/-1, 0/-2, 0/-3, 0/-4, 0/-5}
					{
						\draw[thick] (\i,\j) -- (\i+1,\j);
					}
					\foreach \i/\j in {0/0, 1/0, 0/-1, 1/-1, 0/-2, 1/-2, 0/-3, 1/-3, 0/-4, 1/-4}
					{
						\draw[thick] (\i,\j) -- (\i,\j-1);
					}
					\foreach \i/\j in {}
					{
						\draw[thick] (\i,\j) -- (\i-1,\j-1);
					}
				\end{tikzpicture}
		\end{array}}
	\end{equation}
	
	$6$-th level, row-ordering:
	\begin{equation}
		{\begin{array}{c}
				\begin{tikzpicture}[scale=0.12]
					\foreach \i/\j in {0/0, 1/0, 2/0, 3/0, 4/0, 5/0, 0/-1, 1/-1, 2/-1, 3/-1, 4/-1, 5/-1}
					{
						\draw[thick] (\i,\j) -- (\i+1,\j);
					}
					\foreach \i/\j in {0/0, 1/0, 2/0, 3/0, 4/0, 5/0, 6/0}
					{
						\draw[thick] (\i,\j) -- (\i,\j-1);
					}
					\foreach \i/\j in {}
					{
						\draw[thick] (\i,\j) -- (\i-1,\j-1);
					}
				\end{tikzpicture}
		\end{array}} \succ_{r}
		{\begin{array}{c}
				\begin{tikzpicture}[scale=0.12]
					\foreach \i/\j in {0/0, 1/0, 2/0, 3/0, 4/0, 0/-1, 1/-1, 2/-1, 3/-1, 4/-1, 0/-2}
					{
						\draw[thick] (\i,\j) -- (\i+1,\j);
					}
					\foreach \i/\j in {0/0, 1/0, 2/0, 3/0, 4/0, 0/-1, 1/-1, 5/0}
					{
						\draw[thick] (\i,\j) -- (\i,\j-1);
					}
					\foreach \i/\j in {}
					{
						\draw[thick] (\i,\j) -- (\i-1,\j-1);
					}
				\end{tikzpicture}
		\end{array}} \succ_{r}
		{\begin{array}{c}
				\begin{tikzpicture}[scale=0.12]
					\foreach \i/\j in {0/0, 1/0, 2/0, 3/0, 0/-1, 1/-1, 2/-1, 3/-1, 0/-2, 1/-2}
					{
						\draw[thick] (\i,\j) -- (\i+1,\j);
					}
					\foreach \i/\j in {0/0, 1/0, 2/0, 3/0, 4/0, 0/-1, 1/-1, 2/-1}
					{
						\draw[thick] (\i,\j) -- (\i,\j-1);
					}
					\foreach \i/\j in {}
					{
						\draw[thick] (\i,\j) -- (\i-1,\j-1);
					}
				\end{tikzpicture}
		\end{array}} \succ_{r}
		{\begin{array}{c}
				\begin{tikzpicture}[scale=0.12]
					\foreach \i/\j in {0/0, 1/0, 2/0, 3/0, 0/-1, 1/-1, 2/-1, 3/-1, 0/-2, 0/-3}
					{
						\draw[thick] (\i,\j) -- (\i+1,\j);
					}
					\foreach \i/\j in {0/0, 1/0, 2/0, 3/0, 4/0, 0/-1, 1/-1, 0/-2, 1/-2}
					{
						\draw[thick] (\i,\j) -- (\i,\j-1);
					}
					\foreach \i/\j in {}
					{
						\draw[thick] (\i,\j) -- (\i-1,\j-1);
					}
				\end{tikzpicture}
		\end{array}} \succ_{r}
		{\begin{array}{c}
				\begin{tikzpicture}[scale=0.12]
					\foreach \i/\j in {0/0, 1/0, 2/0, 0/-1, 1/-1, 2/-1, 0/-2, 1/-2, 2/-2}
					{
						\draw[thick] (\i,\j) -- (\i+1,\j);
					}
					\foreach \i/\j in {0/0, 1/0, 2/0, 3/0, 0/-1, 1/-1, 2/-1, 3/-1}
					{
						\draw[thick] (\i,\j) -- (\i,\j-1);
					}
					\foreach \i/\j in {}
					{
						\draw[thick] (\i,\j) -- (\i-1,\j-1);
					}
				\end{tikzpicture}
		\end{array}} \succ_{r}
		{\begin{array}{c}
				\begin{tikzpicture}[scale=0.12]
					\foreach \i/\j in {0/0, 1/0, 2/0, 0/-1, 1/-1, 2/-1, 0/-2, 0/-3, 1/-2}
					{
						\draw[thick] (\i,\j) -- (\i+1,\j);
					}
					\foreach \i/\j in {0/0, 1/0, 2/0, 3/0, 0/-1, 1/-1, 0/-2, 1/-2, 2/-1}
					{
						\draw[thick] (\i,\j) -- (\i,\j-1);
					}
					\foreach \i/\j in {}
					{
						\draw[thick] (\i,\j) -- (\i-1,\j-1);
					}
				\end{tikzpicture}
		\end{array}} \succ_{r}
		{\begin{array}{c}
				\begin{tikzpicture}[scale=0.12]
					\foreach \i/\j in {0/0, 1/0, 2/0, 0/-1, 1/-1, 2/-1, 0/-2, 0/-3, 0/-4}
					{
						\draw[thick] (\i,\j) -- (\i+1,\j);
					}
					\foreach \i/\j in {0/0, 1/0, 2/0, 3/0, 0/-1, 1/-1, 0/-2, 1/-2, 0/-3, 1/-3}
					{
						\draw[thick] (\i,\j) -- (\i,\j-1);
					}
					\foreach \i/\j in {}
					{
						\draw[thick] (\i,\j) -- (\i-1,\j-1);
					}
				\end{tikzpicture}
		\end{array}} \succ_{r}
		{\begin{array}{c}
				\begin{tikzpicture}[scale=0.12]
					\foreach \i/\j in {0/0, 1/0, 0/-1, 1/-1, 0/-2, 0/-3, 1/-2, 1/-3}
					{
						\draw[thick] (\i,\j) -- (\i+1,\j);
					}
					\foreach \i/\j in {0/0, 1/0, 2/0, 0/-1, 1/-1, 0/-2, 1/-2, 2/-1, 2/-2}
					{
						\draw[thick] (\i,\j) -- (\i,\j-1);
					}
					\foreach \i/\j in {}
					{
						\draw[thick] (\i,\j) -- (\i-1,\j-1);
					}
				\end{tikzpicture}
		\end{array}} \succ_{r}
		{\begin{array}{c}
				\begin{tikzpicture}[scale=0.12]
					\foreach \i/\j in {0/0, 1/0, 0/-1, 1/-1, 0/-2, 0/-3, 1/-2, 0/-4}
					{
						\draw[thick] (\i,\j) -- (\i+1,\j);
					}
					\foreach \i/\j in {0/0, 1/0, 2/0, 0/-1, 1/-1, 0/-2, 1/-2, 2/-1, 0/-3, 1/-3}
					{
						\draw[thick] (\i,\j) -- (\i,\j-1);
					}
					\foreach \i/\j in {}
					{
						\draw[thick] (\i,\j) -- (\i-1,\j-1);
					}
				\end{tikzpicture}
		\end{array}} \succ_{r}
		{\begin{array}{c}
				\begin{tikzpicture}[scale=0.12]
					\foreach \i/\j in {0/0, 1/0, 0/-1, 1/-1, 0/-2, 0/-3, 0/-4, 0/-5}
					{
						\draw[thick] (\i,\j) -- (\i+1,\j);
					}
					\foreach \i/\j in {0/0, 1/0, 2/0, 0/-1, 1/-1, 0/-2, 1/-2, 0/-3, 1/-3, 0/-4, 1/-4}
					{
						\draw[thick] (\i,\j) -- (\i,\j-1);
					}
					\foreach \i/\j in {}
					{
						\draw[thick] (\i,\j) -- (\i-1,\j-1);
					}
				\end{tikzpicture}
		\end{array}} \succ_{r}
		{\begin{array}{c}
				\begin{tikzpicture}[scale=0.12]
					\foreach \i/\j in {0/0, 0/-1, 0/-2, 0/-3, 0/-4, 0/-5, 0/-6}
					{
						\draw[thick] (\i,\j) -- (\i+1,\j);
					}
					\foreach \i/\j in {0/0, 1/0, 0/-1, 1/-1, 0/-2, 1/-2, 0/-3, 1/-3, 0/-4, 1/-4, 0/-5, 1/-5}
					{
						\draw[thick] (\i,\j) -- (\i,\j-1);
					}
					\foreach \i/\j in {}
					{
						\draw[thick] (\i,\j) -- (\i-1,\j-1);
					}
				\end{tikzpicture}
		\end{array}}
	\end{equation}
	
	$6$-th level, column-ordering:
	\begin{equation}
		{\begin{array}{c}
				\begin{tikzpicture}[scale=0.12]
					\foreach \i/\j in {0/0, 1/0, 2/0, 3/0, 4/0, 5/0, 0/-1, 1/-1, 2/-1, 3/-1, 4/-1, 5/-1}
					{
						\draw[thick] (\i,\j) -- (\i+1,\j);
					}
					\foreach \i/\j in {0/0, 1/0, 2/0, 3/0, 4/0, 5/0, 6/0}
					{
						\draw[thick] (\i,\j) -- (\i,\j-1);
					}
					\foreach \i/\j in {}
					{
						\draw[thick] (\i,\j) -- (\i-1,\j-1);
					}
				\end{tikzpicture}
		\end{array}} \succ_{c}
		{\begin{array}{c}
				\begin{tikzpicture}[scale=0.12]
					\foreach \i/\j in {0/0, 1/0, 2/0, 3/0, 4/0, 0/-1, 1/-1, 2/-1, 3/-1, 4/-1, 0/-2}
					{
						\draw[thick] (\i,\j) -- (\i+1,\j);
					}
					\foreach \i/\j in {0/0, 1/0, 2/0, 3/0, 4/0, 0/-1, 1/-1, 5/0}
					{
						\draw[thick] (\i,\j) -- (\i,\j-1);
					}
					\foreach \i/\j in {}
					{
						\draw[thick] (\i,\j) -- (\i-1,\j-1);
					}
				\end{tikzpicture}
		\end{array}} \succ_{c}
		{\begin{array}{c}
				\begin{tikzpicture}[scale=0.12]
					\foreach \i/\j in {0/0, 1/0, 2/0, 3/0, 0/-1, 1/-1, 2/-1, 3/-1, 0/-2, 1/-2}
					{
						\draw[thick] (\i,\j) -- (\i+1,\j);
					}
					\foreach \i/\j in {0/0, 1/0, 2/0, 3/0, 4/0, 0/-1, 1/-1, 2/-1}
					{
						\draw[thick] (\i,\j) -- (\i,\j-1);
					}
					\foreach \i/\j in {}
					{
						\draw[thick] (\i,\j) -- (\i-1,\j-1);
					}
				\end{tikzpicture}
		\end{array}} \succ_{c}
		\underline{\begin{array}{c}
				\begin{tikzpicture}[scale=0.12]
					\foreach \i/\j in {0/0, 1/0, 2/0, 0/-1, 1/-1, 2/-1, 0/-2, 1/-2, 2/-2}
					{
						\draw[thick] (\i,\j) -- (\i+1,\j);
					}
					\foreach \i/\j in {0/0, 1/0, 2/0, 3/0, 0/-1, 1/-1, 2/-1, 3/-1}
					{
						\draw[thick] (\i,\j) -- (\i,\j-1);
					}
					\foreach \i/\j in {}
					{
						\draw[thick] (\i,\j) -- (\i-1,\j-1);
					}
				\end{tikzpicture}
		\end{array}} \succ_{c}
		\underline{\begin{array}{c}
				\begin{tikzpicture}[scale=0.12]
					\foreach \i/\j in {0/0, 1/0, 2/0, 3/0, 0/-1, 1/-1, 2/-1, 3/-1, 0/-2, 0/-3}
					{
						\draw[thick] (\i,\j) -- (\i+1,\j);
					}
					\foreach \i/\j in {0/0, 1/0, 2/0, 3/0, 4/0, 0/-1, 1/-1, 0/-2, 1/-2}
					{
						\draw[thick] (\i,\j) -- (\i,\j-1);
					}
					\foreach \i/\j in {}
					{
						\draw[thick] (\i,\j) -- (\i-1,\j-1);
					}
				\end{tikzpicture}
		\end{array}} \succ_{c}
		{\begin{array}{c}
				\begin{tikzpicture}[scale=0.12]
					\foreach \i/\j in {0/0, 1/0, 2/0, 0/-1, 1/-1, 2/-1, 0/-2, 0/-3, 1/-2}
					{
						\draw[thick] (\i,\j) -- (\i+1,\j);
					}
					\foreach \i/\j in {0/0, 1/0, 2/0, 3/0, 0/-1, 1/-1, 0/-2, 1/-2, 2/-1}
					{
						\draw[thick] (\i,\j) -- (\i,\j-1);
					}
					\foreach \i/\j in {}
					{
						\draw[thick] (\i,\j) -- (\i-1,\j-1);
					}
				\end{tikzpicture}
		\end{array}} \succ_{c}
		\underline{\begin{array}{c}
				\begin{tikzpicture}[scale=0.12]
					\foreach \i/\j in {0/0, 1/0, 0/-1, 1/-1, 0/-2, 0/-3, 1/-2, 1/-3}
					{
						\draw[thick] (\i,\j) -- (\i+1,\j);
					}
					\foreach \i/\j in {0/0, 1/0, 2/0, 0/-1, 1/-1, 0/-2, 1/-2, 2/-1, 2/-2}
					{
						\draw[thick] (\i,\j) -- (\i,\j-1);
					}
					\foreach \i/\j in {}
					{
						\draw[thick] (\i,\j) -- (\i-1,\j-1);
					}
				\end{tikzpicture}
		\end{array}} \succ_{c}
		\underline{\begin{array}{c}
				\begin{tikzpicture}[scale=0.12]
					\foreach \i/\j in {0/0, 1/0, 2/0, 0/-1, 1/-1, 2/-1, 0/-2, 0/-3, 0/-4}
					{
						\draw[thick] (\i,\j) -- (\i+1,\j);
					}
					\foreach \i/\j in {0/0, 1/0, 2/0, 3/0, 0/-1, 1/-1, 0/-2, 1/-2, 0/-3, 1/-3}
					{
						\draw[thick] (\i,\j) -- (\i,\j-1);
					}
					\foreach \i/\j in {}
					{
						\draw[thick] (\i,\j) -- (\i-1,\j-1);
					}
				\end{tikzpicture}
		\end{array}} \succ_{c}
		{\begin{array}{c}
				\begin{tikzpicture}[scale=0.12]
					\foreach \i/\j in {0/0, 1/0, 0/-1, 1/-1, 0/-2, 0/-3, 1/-2, 0/-4}
					{
						\draw[thick] (\i,\j) -- (\i+1,\j);
					}
					\foreach \i/\j in {0/0, 1/0, 2/0, 0/-1, 1/-1, 0/-2, 1/-2, 2/-1, 0/-3, 1/-3}
					{
						\draw[thick] (\i,\j) -- (\i,\j-1);
					}
					\foreach \i/\j in {}
					{
						\draw[thick] (\i,\j) -- (\i-1,\j-1);
					}
				\end{tikzpicture}
		\end{array}} \succ_{c}
		{\begin{array}{c}
				\begin{tikzpicture}[scale=0.12]
					\foreach \i/\j in {0/0, 1/0, 0/-1, 1/-1, 0/-2, 0/-3, 0/-4, 0/-5}
					{
						\draw[thick] (\i,\j) -- (\i+1,\j);
					}
					\foreach \i/\j in {0/0, 1/0, 2/0, 0/-1, 1/-1, 0/-2, 1/-2, 0/-3, 1/-3, 0/-4, 1/-4}
					{
						\draw[thick] (\i,\j) -- (\i,\j-1);
					}
					\foreach \i/\j in {}
					{
						\draw[thick] (\i,\j) -- (\i-1,\j-1);
					}
				\end{tikzpicture}
		\end{array}} \succ_{c}
		{\begin{array}{c}
				\begin{tikzpicture}[scale=0.12]
					\foreach \i/\j in {0/0, 0/-1, 0/-2, 0/-3, 0/-4, 0/-5, 0/-6}
					{
						\draw[thick] (\i,\j) -- (\i+1,\j);
					}
					\foreach \i/\j in {0/0, 1/0, 0/-1, 1/-1, 0/-2, 1/-2, 0/-3, 1/-3, 0/-4, 1/-4, 0/-5, 1/-5}
					{
						\draw[thick] (\i,\j) -- (\i,\j-1);
					}
					\foreach \i/\j in {}
					{
						\draw[thick] (\i,\j) -- (\i-1,\j-1);
					}
				\end{tikzpicture}
		\end{array}}
	\end{equation}
	
	Remarkably, triangular decomposition respects by the second (column) ordering too:
	\begin{tcolorbox}
		\begin{equation}
			\label{triangularity Macdonald}
			M_{\lambda}^{q,t} = S_{\lambda} + \sum_{\mu \, \prec_{c} \, \lambda} K_{\lambda, \mu}^{q,t} \, S_{\mu}
		\end{equation}
	\end{tcolorbox}
	Note that Macdonald-Kostka coefficients of this column-decomposition is the same as in row-decomposition formulas \eqref{row triangularity Macdonald} because Schur polynomials form a basis and basis coefficients are always unique. This can happen only if "uncertain" Macdonald-Kostka coefficients vanish. For our example from $6$-th level it means:
	\begin{equation}
		K^{q,t}_{[4,1,1], [3,3]} = 0, \hspace{10mm} K^{q,t}_{[3,3], [4,1,1]} = 0
	\end{equation}
	\begin{equation}
		K^{q,t}_{[3,1,1,1], [2,2,2]} = 0, \hspace{10mm} K^{q,t}_{[2,2,2], [3,1,1,1]} = 0
	\end{equation}
	
	The fact that the Macdonald polynomials admit two different interpretation in terms of row- and column- triangular decomposition in Schur basis is a nontrivial property of Macdonald polynomials. In particular, this phenomenon is a property of Macdonald $q,t$-measure \eqref{q,t measure Macdonald} and naive deviation of it (\textit{a la} Kerov polynomials \cite{kerov1991hall}) immediately destroys the phenomenon \cite{Mironov:2020aaa}.

	$\bullet$ \underline{Transposing relations}. Alternatively Cauchy relation my be represented in the following form:
	\begin{equation}
		\label{alternative Cauchy for Macdonald}
		\sum_{\lambda} (-)^{|\lambda|} \, M^{\, q,t}_{\lambda}(p) \cdot M^{\, t,q}_{\lambda^T}(-\bar{p}) = \exp\left( \sum_{n=1} \frac{p_n \, \bar{p}_n}{n}  \right)
	\end{equation}
	Note that in the second Macdonald polynomial the Young diagram is transposed $\lambda \to \lambda^T$ and $q,t$ parameters are interchanged $q \leftrightarrow t$. The above relation allows to deduce peculiar transposing formula for Macdonald polynomials:
	\begin{tcolorbox}
		\begin{equation}
			\label{transposing rule}
			(-)^{|\lambda|} \cdot \frac{M^{q,t}_{\lambda}(p_k)}{|| M^{q,t}_{\lambda} ||^2} =   M^{\, t,q}_{\lambda^T}\left( (-) \cdot \frac{t^{2k} - 1}{q^{2k} - 1} \cdot p_k \right)
		\end{equation}
	\end{tcolorbox}
	These formulas impose conditions on normalizations of Macdonald polynomials $|| M^{q,t}_{\lambda}||$.

\section{Super-Schur polynomials from $Y(\hat{\mathfrak{gl}}_{1|1})$} \label{sec:super-Schur}

One of the main ingredients of super-Macdonald $\mathcal{M}^{q,t}_{\lambda}(p_k, \theta_k)$ polynomials is the proper generalization of Schur polynomials $S_{\lambda}(p_k)$ to the case of super-partitions. Suitable family of polynomials, that we call \underline{\textit{ super-Schur polynomials} $\mathcal{S}_{\lambda}(p_k, \theta_k)$}, arises in a free-field realization of semi-Fock representation of affine super-Yangian $Y(\hat{\mathfrak{gl}}_{1|1})$ \cite{Galakhov:2023mak}, where vectors in the representation are in one-to-one correspondence with super-partitions. Super-Schur polynomials are polynomials in double set of variables: usual $p_k$ and Grassmann $\theta_k$ variables, where $k \in \mathbb{N}$. Alike usual Schur polynomials super-Schurs do not have any deformation parameters \footnote{In \cite{} we used the name "super-Schur polynomials" for polynomials that depend on two deformation parameters $\epsilon_1, \epsilon_2$ and therefore are relatives of Jack polynomials with $\beta$-deformation parameter, where $\beta = - \epsilon_2/ \epsilon_1$. Here we use slightly different notation, super-Schurs in this paper do not depend on parameters and correspond to the case $\epsilon_1 = 1, \epsilon_2 = -1$.}.

Following the same strategy as for usual Schur polynomials, we define here super-Schur polynomials as common eigenfunctions of commuting set of operators $\hat{\psi}^{+}_n$ and $\hat{\psi}^{-}_n$, where $n \in \mathbb{N}_0$:

\begin{equation}
	\hat{\psi}^{\pm}_n \, \mathcal{S}_{\lambda}(p_k, \theta_k) = \mathcal{E}^{\pm}_{n,\lambda} \, \mathcal{S}_{\lambda}(p_k, \theta_k)
\end{equation}

These operators form Cartan subalgebra of $Y(\hat{\mathfrak{gl}}_{1|1})$ and we construct them in four steps:
\begin{enumerate}
	\item On the first step we define six auxiliary operators:
	\begin{align}
		\begin{aligned}
			\hat{e}_0^{+} &:= \theta_1, &\hspace{20mm} \hat{f}_0^{+} &:= \frac{\partial}{\partial \theta_1} \\
			\hat{e}_0^{-} &:= \sum_{k=1} p_k \frac{\partial}{\partial \theta_k}, &\hspace{20mm} \hat{f}_0^{-} &:= -\sum_{k=1} k \theta_k \frac{\partial}{\partial p_k}
		\end{aligned}
	\end{align}
	\begin{equation}
		\begin{aligned}
			\hat{W}^{\pm}&:=\frac{1}{2}\sum_{a,b= 1}^{\infty}ab \,p_{a+b}\frac{\partial^2}{\partial p_a \partial p_b}+(a+b)p_ap_b\frac{\partial}{\partial p_{a+b}}
			+\\
			&+\sum_{a,b=1}^{\infty}\left(b - \frac{1\mp1}{2} \right)\cdot\left[a\,\theta_{a+b}\frac{\partial^2}{\partial p_a\partial\theta_b}+p_a\theta_b\frac{\partial}{\partial\theta_{a+b}}\right]
		\end{aligned}
	\end{equation}

	\item On the second step we compute higher auxiliary operators recursively by the following formula:
	\begin{align}
		\hat{e}^{\pm}_{n+1} &:= \Big[ \hat{W}^{\pm}, \hat{e}^{\pm}_{n}\Big]
	\end{align}
	\item On the third step we compute operators $\hat{\psi}_{n}^{\pm}$ as anti-commutators:
	\begin{align}
		\hat{\psi}^{\pm}_{n} := \Big\{ \hat{e}_{n}^{\pm}, \hat{f}_{0}^{\pm} \Big\}
	\end{align}
	The resulting operators form a commuting set:
	\begin{align}
		\Big[ \hat{\psi}^{\pm}_n, \hat{\psi}^{\pm}_m \Big] = 0 \hspace{25mm} \Big[ \hat{\psi}^{+}_n, \hat{\psi}^{-}_m \Big] = 0
	\end{align}
	In particular, the initial operators $\hat{W}^{\pm}$ are also included in this set:
	\begin{align}
		\begin{aligned}
			\hat{W}^{+} = -\frac{1}{2} \hat{\psi}^{-}_1 \hspace{25mm} \hat{W}^{-} = \frac{1}{2} \hat{\psi}^{+}_3
		\end{aligned}
	\end{align}
	\item On the final step we compute eigenvalues $\mathcal{E}^{\pm}_{n,\lambda}$ in order to identify the super-Schur polynomials. These eigenvalues follows from the representation theory of the affine super-Yangian \cite{Galakhov:2023mak, Galakhov:2024foa}. The eigenvalues are collected in the generating functions:
	\begin{equation}
		\mathcal{E}^{+}_{\lambda} (z) =  \sum_{n = 0} \frac{\mathcal{E}^{+}_{n,\lambda}}{z^{n+1}} \hspace{10mm} \mathcal{E}^{-}_{\lambda} (z) =  z + \sum_{n = 0} \frac{\mathcal{E}^{-}_{n,\lambda}}{z^{n+1}}
	\end{equation}
	The generating functions itself are given by the following products over boxes and half-boxes of the super-Young diagrams:
	\begin{equation}
		\mathcal{E}^{+}_{\lambda} (z) = \frac{1}{z} \cdot \prod_{\begin{tikzpicture}[scale=0.15]
				\foreach \i/\j in {0/0, 0/-1}
				{
					\draw[thick] (\i,\j) -- (\i+1,\j);
				}
				\foreach \i/\j in {0/0, 1/0}
				{
					\draw[thick] (\i,\j) -- (\i,\j-1);
				}
				\foreach \i/\j in {}
				{
					\draw[thick] (\i,\j) -- (\i-1,\j-1);
				}
		\end{tikzpicture} \, \in \lambda} \phi \left(z - \omega_{\begin{tikzpicture}[scale=0.15]
		\foreach \i/\j in {0/0, 0/-1}
		{
			\draw[thick] (\i,\j) -- (\i+1,\j);
		}
		\foreach \i/\j in {0/0, 1/0}
		{
			\draw[thick] (\i,\j) -- (\i,\j-1);
		}
		\foreach \i/\j in {}
		{
			\draw[thick] (\i,\j) -- (\i-1,\j-1);
		}
	\end{tikzpicture}} \right) \hspace{20mm}
		\mathcal{E}^{-}_{\lambda} (z) = z \cdot \prod_{\begin{tikzpicture}[scale=0.15]
				\foreach \i/\j in {0/0, 0/-1}
				{
					\draw[thick] (\i,\j) -- (\i+1,\j);
				}
				\foreach \i/\j in {0/0, 1/0}
				{
					\draw[thick] (\i,\j) -- (\i,\j-1);
				}
				\foreach \i/\j in {}
				{
					\draw[thick] (\i,\j) -- (\i-1,\j-1);
				}
			\end{tikzpicture} \, \in \lambda} \frac{1}{\phi\left( z - \omega_{\begin{tikzpicture}[scale=0.15]
				\foreach \i/\j in {0/0, 0/-1}
				{
					\draw[thick] (\i,\j) -- (\i+1,\j);
				}
				\foreach \i/\j in {0/0, 1/0}
				{
					\draw[thick] (\i,\j) -- (\i,\j-1);
				}
				\foreach \i/\j in {}
				{
					\draw[thick] (\i,\j) -- (\i-1,\j-1);
				}
		\end{tikzpicture}}\right)}
		\cdot \prod_{\begin{tikzpicture}[scale=0.15]
				\foreach \i/\j in {0/0}
				{
					\draw[thick] (\i,\j) -- (\i+1,\j);
				}
				\foreach \i/\j in {0/0}
				{
					\draw[thick] (\i,\j) -- (\i,\j-1);
				}
				\foreach \i/\j in {1/0}
				{
					\draw[thick] (\i,\j) -- (\i-1,\j-1);
				}
			\end{tikzpicture} \, \in \lambda} \frac{1}{\phi\left( z - \omega_{\begin{tikzpicture}[scale=0.15]
					\foreach \i/\j in {0/0}
					{
						\draw[thick] (\i,\j) -- (\i+1,\j);
					}
					\foreach \i/\j in {0/0}
					{
						\draw[thick] (\i,\j) -- (\i,\j-1);
					}
					\foreach \i/\j in {1/0}
					{
						\draw[thick] (\i,\j) -- (\i-1,\j-1);
					}
			\end{tikzpicture}}\right)}
	\end{equation}
	Here the function $\phi(z)$ is defined as follows:
	\begin{equation}
		\phi(z) = \frac{z^2}{z^2 - 1}
	\end{equation}
	Boxes and half-boxes have vertical and horizontal coordinates $(i,j)$ and content is defined as $\omega := i - j$. In order to show our notations we provide an example:
	\begin{align}
		\begin{aligned}
			\mathcal{E}^{+}_{\, \begin{tikzpicture}[scale=0.15]
					\foreach \i/\j in {0/0, 0/-1, 1/0, 1/-1, 2/0}
					{
						\draw[thick] (\i,\j) -- (\i+1,\j);
					}
					\foreach \i/\j in {0/0, 1/0, 2/0, 0/-1}
					{
						\draw[thick] (\i,\j) -- (\i,\j-1);
					}
					\foreach \i/\j in {3/0, 1/-1}
					{
						\draw[thick] (\i,\j) -- (\i-1,\j-1);
					}
			\end{tikzpicture}} (z) &= \frac{\phi(z) \phi(z+1)}{z} = \frac{1}{z}+\frac{2}{z^3}-\frac{2}{z^4}+\frac{6}{z^5}-\frac{10}{z^6} + \ldots \\
			\mathcal{E}^{-}_{\, \begin{tikzpicture}[scale=0.15]
					\foreach \i/\j in {0/0, 0/-1, 1/0, 1/-1, 2/0}
					{
						\draw[thick] (\i,\j) -- (\i+1,\j);
					}
					\foreach \i/\j in {0/0, 1/0, 2/0, 0/-1}
					{
						\draw[thick] (\i,\j) -- (\i,\j-1);
					}
					\foreach \i/\j in {3/0, 1/-1}
					{
						\draw[thick] (\i,\j) -- (\i-1,\j-1);
					}
			\end{tikzpicture}} (z) &= \frac{z}{\underbrace{\phi(z) \phi(z + 1)}_{\text{Boxes}} \cdot \underbrace{\phi(z - 1) \phi(z+2)}_{\text{Half-boxes}}} = z-\frac{4}{z}+\frac{4}{z^2}-\frac{12}{z^3}+\frac{20}{z^4}-\frac{44}{z^5}+\frac{84}{z^6}+ \ldots
		\end{aligned}
	\end{align}
\end{enumerate}
The above defined explicit operators $\hat{\psi}_{n}^{\pm}$ define a set of common eigenfunctions:
\begin{equation}
	\psi^{\pm}_n \, \mathcal{S}_{\lambda}(p, \theta) = \mathcal{E}^{\pm}_{n,\lambda} \, \mathcal{S}_{\lambda}(p, \theta)
\end{equation}
The above definition determines the form of super-Schur polynomials up to overall scaling factor. We fix this factor by the following orthonormal requirement:
\begin{equation}
	\bra{\mathcal{S}_{\lambda}} \ket{\mathcal{S}_{\mu} } = \delta_{\lambda, \mu}
\end{equation}
where we generalized the scalar product by the following rule:
\begin{equation}
	\label{scalar product super-Schur}
	\bra{p_{\Delta} \theta_{\Lambda}} \ket{p_{\Delta^{\prime}} \theta_{\Lambda^{\prime}}} = \delta_{\Delta, \Delta^{\prime}} \cdot \delta_{\Lambda, \Lambda^{\prime}} \cdot z_{\Delta}
\end{equation}
In this formula $\Lambda = \left[ \Lambda_1, \Lambda_2, \ldots, \Lambda_{l(\Lambda)} \right]$ is a strict Young diagram, i.e. $\Lambda_1 > \Lambda_2 > \ldots > \Lambda_{l(\Lambda)}$. Grassmann monomials $\theta_{\Lambda}$ are defined as follows:
\begin{equation}
	\theta_{\Lambda} := \theta_{\Lambda_{l(\Lambda)}} \ldots \theta_{\Lambda_2} \, \theta_{\Lambda_1}
\end{equation}
Note that the order of Grassmann variables determines the overall sign and we order the variables in a way that the indices increase from left to right. Also note that Grassmann variables have trivial norm on the r.h.s. of \eqref{scalar product super-Schur}.
Generalized Cauchy identity reflects the same orthogonality property:
\begin{equation}
	\sum_{\lambda} \, \mathcal{S}_{\lambda} (p,\theta) \cdot \mathcal{S}_{\lambda}(\bar{p},\bar{\theta}) = \exp\left( \sum_{n=1} \frac{p_n \, \bar{p}_n}{n} + \theta_n \, \bar{\theta}_n \right)
\end{equation}
In this formula the sum over diagrams $\lambda$ runs over all super-Young diagrams. We list explicit examples of super-Schur polynomials from small levels:

\begin{align}
	\begin{aligned}
		\mathcal{S}_{\, \varnothing} &= 1 &
		\hspace{10mm}
		\mathcal{S}_{\, \begin{tikzpicture}[scale=0.15]
				\foreach \i/\j in {0/0, 0/-1, 1/0, 1/-1, 2/0}
				{
					\draw[thick] (\i,\j) -- (\i+1,\j);
				}
				\foreach \i/\j in {0/0, 1/0, 2/0}
				{
					\draw[thick] (\i,\j) -- (\i,\j-1);
				}
				\foreach \i/\j in {3/0}
				{
					\draw[thick] (\i,\j) -- (\i-1,\j-1);
				}
		\end{tikzpicture}} &= \frac{1}{\sqrt{3}} \left( \frac{p_1^2 \theta_1}{2} + \frac{p_2 \theta_1}{2} + p_1 \theta_2 + \theta_3 \right)\\
		\mathcal{S}_{\, \begin{tikzpicture}[scale=0.15]
				\foreach \i/\j in {0/0}
				{
					\draw[thick] (\i,\j) -- (\i+1,\j);
				}
				\foreach \i/\j in {0/0}
				{
					\draw[thick] (\i,\j) -- (\i,\j-1);
				}
				\foreach \i/\j in {1/0}
				{
					\draw[thick] (\i,\j) -- (\i-1,\j-1);
				}
		\end{tikzpicture}} &= \theta_1 &
		\hspace{10mm}
		\mathcal{S}_{\, \begin{tikzpicture}[scale=0.15]
				\foreach \i/\j in {0/0, 0/-1, 1/0, 1/-1}
				{
					\draw[thick] (\i,\j) -- (\i+1,\j);
				}
				\foreach \i/\j in {0/0, 1/0, 0/-1, 2/0}
				{
					\draw[thick] (\i,\j) -- (\i,\j-1);
				}
				\foreach \i/\j in {1/-1}
				{
					\draw[thick] (\i,\j) -- (\i-1,\j-1);
				}
		\end{tikzpicture}} &= \frac{1}{\sqrt{6}} \left( p_1^2 \theta_1 + p_2 \theta_1 - p_1 \theta_2 - \theta_3 \right) \\
		\mathcal{S}_{\, \begin{tikzpicture}[scale=0.15]
				\foreach \i/\j in {0/0, 0/-1}
				{
					\draw[thick] (\i,\j) -- (\i+1,\j);
				}
				\foreach \i/\j in {0/0, 1/0}
				{
					\draw[thick] (\i,\j) -- (\i,\j-1);
				}
				\foreach \i/\j in {}
				{
					\draw[thick] (\i,\j) -- (\i-1,\j-1);
				}
		\end{tikzpicture}} &= p_1 &
		\hspace{10mm}
		\mathcal{S}_{\, \begin{tikzpicture}[scale=0.15]
				\foreach \i/\j in {0/0, 0/-1, 1/0, 0/-2}
				{
					\draw[thick] (\i,\j) -- (\i+1,\j);
				}
				\foreach \i/\j in {0/0, 1/0, 0/-1, 1/-1}
				{
					\draw[thick] (\i,\j) -- (\i,\j-1);
				}
				\foreach \i/\j in {2/0}
				{
					\draw[thick] (\i,\j) -- (\i-1,\j-1);
				}
		\end{tikzpicture}} &= \frac{1}{\sqrt{6}} \left( p_1^2 \theta_1 - p_2 \theta_1 + p_1 \theta_2 - \theta_3 \right) \\
		\mathcal{S}_{\, \begin{tikzpicture}[scale=0.15]
				\foreach \i/\j in {0/0, 0/-1, 1/0}
				{
					\draw[thick] (\i,\j) -- (\i+1,\j);
				}
				\foreach \i/\j in {0/0, 1/0}
				{
					\draw[thick] (\i,\j) -- (\i,\j-1);
				}
				\foreach \i/\j in {2/0}
				{
					\draw[thick] (\i,\j) -- (\i-1,\j-1);
				}
		\end{tikzpicture}} &= \frac{1}{\sqrt{2}} \left( p_1 \theta_1 + \theta_2 \right) &
		\hspace{10mm}
		\mathcal{S}_{\, \begin{tikzpicture}[scale=0.15]
				\foreach \i/\j in {0/0, 0/-1, 0/-2}
				{
					\draw[thick] (\i,\j) -- (\i+1,\j);
				}
				\foreach \i/\j in {0/0, 1/0, 0/-1, 1/-1, 0/-2}
				{
					\draw[thick] (\i,\j) -- (\i,\j-1);
				}
				\foreach \i/\j in {1/-2}
				{
					\draw[thick] (\i,\j) -- (\i-1,\j-1);
				}
		\end{tikzpicture} } &= \frac{1}{\sqrt{3}} \left( \frac{p_1^2 \theta_1}{2} - \frac{p_2 \theta_1}{2} - p_1 \theta_2 + \theta_3 \right) \\
		\mathcal{S}_{\, \begin{tikzpicture}[scale=0.15]
				\foreach \i/\j in {0/0, 0/-1}
				{
					\draw[thick] (\i,\j) -- (\i+1,\j);
				}
				\foreach \i/\j in {0/0, 1/0, 0/-1}
				{
					\draw[thick] (\i,\j) -- (\i,\j-1);
				}
				\foreach \i/\j in {1/-1}
				{
					\draw[thick] (\i,\j) -- (\i-1,\j-1);
				}
		\end{tikzpicture}} &= \frac{1}{\sqrt{2}} \left( p_1 \theta_1 - \theta_2 \right) &
		\hspace{10mm}
		\mathcal{S}_{\, \begin{tikzpicture}[scale=0.15]
				\foreach \i/\j in {0/0, 0/-1, 1/0, 1/-1, 2/0, 2/-1}
				{
					\draw[thick] (\i,\j) -- (\i+1,\j);
				}
				\foreach \i/\j in {0/0, 1/0, 2/0, 3/0}
				{
					\draw[thick] (\i,\j) -- (\i,\j-1);
				}
				\foreach \i/\j in {}
				{
					\draw[thick] (\i,\j) -- (\i-1,\j-1);
				}
		\end{tikzpicture}} &= \frac{p_1^3}{6} + \frac{p_1 p_2}{2} + \frac{p_3}{3}\\
		\mathcal{S}_{\, \begin{tikzpicture}[scale=0.15]
				\foreach \i/\j in {0/0, 0/-1, 1/0, 1/-1}
				{
					\draw[thick] (\i,\j) -- (\i+1,\j);
				}
				\foreach \i/\j in {0/0, 1/0, 2/0}
				{
					\draw[thick] (\i,\j) -- (\i,\j-1);
				}
				\foreach \i/\j in {}
				{
					\draw[thick] (\i,\j) -- (\i-1,\j-1);
				}
			\end{tikzpicture}} &= \frac{p_1^2}{2} + \frac{p_2}{2} &
		\hspace{10mm}
		\mathcal{S}_{\,\begin{tikzpicture}[scale=0.15]
				\foreach \i/\j in {0/0, 0/-1, 0/-2, 1/0, 1/-1}
				{
					\draw[thick] (\i,\j) -- (\i+1,\j);
				}
				\foreach \i/\j in {0/0, 1/0, 0/-1, 1/-1, 2/0}
				{
					\draw[thick] (\i,\j) -- (\i,\j-1);
				}
				\foreach \i/\j in {}
				{
					\draw[thick] (\i,\j) -- (\i-1,\j-1);
				}
		\end{tikzpicture}} &= \frac{p_1^3}{3} - \frac{p_3}{3}\\
		\mathcal{S}_{\, \begin{tikzpicture}[scale=0.15]
			\foreach \i/\j in {0/0, 0/-1, 0/-2}
			{
				\draw[thick] (\i,\j) -- (\i+1,\j);
			}
			\foreach \i/\j in {0/0, 1/0, 0/-1, 1/-1}
			{
				\draw[thick] (\i,\j) -- (\i,\j-1);
			}
			\foreach \i/\j in {}
			{
				\draw[thick] (\i,\j) -- (\i-1,\j-1);
			}
		\end{tikzpicture} }  &= \frac{p_1^2}{2} - \frac{p_2}{2} &
		\hspace{10mm}
		\mathcal{S}_{\, \begin{tikzpicture}[scale=0.15]
				\foreach \i/\j in {0/0, 0/-1, 0/-2, 0/-3}
				{
					\draw[thick] (\i,\j) -- (\i+1,\j);
				}
				\foreach \i/\j in {0/0, 1/0, 0/-1, 1/-1, 0/-2, 1/-2}
				{
					\draw[thick] (\i,\j) -- (\i,\j-1);
				}
				\foreach \i/\j in {}
				{
					\draw[thick] (\i,\j) -- (\i-1,\j-1);
				}
		\end{tikzpicture}} &= \frac{p_1^3}{6} - \frac{p_1 p_2}{2} + \frac{p_3}{3} \\
		\mathcal{S}_{\, \begin{tikzpicture}[scale=0.15]
		\foreach \i/\j in {0/0, 0/-1, 1/0}
		{
			\draw[thick] (\i,\j) -- (\i+1,\j);
		}
		\foreach \i/\j in {0/0, 1/0, 0/-1}
		{
			\draw[thick] (\i,\j) -- (\i,\j-1);
		}
		\foreach \i/\j in {2/0, 1/-1}
		{
			\draw[thick] (\i,\j) -- (\i-1,\j-1);
		} \end{tikzpicture}} &= \theta_1 \theta_2 &
		\hspace{10mm}
		\mathcal{S}_{\, \begin{tikzpicture}[scale=0.15]
				\foreach \i/\j in {0/0, 0/-1, 1/0, 1/-1, 2/0}
				{
					\draw[thick] (\i,\j) -- (\i+1,\j);
				}
				\foreach \i/\j in {0/0, 1/0, 2/0, 0/-1}
				{
					\draw[thick] (\i,\j) -- (\i,\j-1);
				}
				\foreach \i/\j in {3/0, 1/-1}
				{
					\draw[thick] (\i,\j) -- (\i-1,\j-1);
				}
		\end{tikzpicture}} &= \frac{1}{\sqrt{2}} \left( p_1 \theta_1 \theta_2 + \theta_1 \theta_3 \right) \\
		&  &\hspace{10mm} \mathcal{S}_{\, \begin{tikzpicture}[scale=0.15]
				\foreach \i/\j in {0/0, 0/-1, 1/0, 0/-2}
				{
					\draw[thick] (\i,\j) -- (\i+1,\j);
				}
				\foreach \i/\j in {0/0, 1/0, 0/-1, 1/-1, 0/-2}
				{
					\draw[thick] (\i,\j) -- (\i,\j-1);
				}
				\foreach \i/\j in {2/0, 1/-2}
				{
					\draw[thick] (\i,\j) -- (\i-1,\j-1);
				}
		\end{tikzpicture}} &= \frac{1}{\sqrt{2}} \left( p_1 \theta_1 \theta_2 - \theta_1 \theta_3 \right) \\
	\end{aligned}
\end{align}
For usual Young diagrams, i.e. super-Young diagrams without triangles, super-Schur polynomials coincide with Schur polynomials:
\begin{equation}
	\label{Schur = super-Schur}
	\mathcal{S}_{\lambda} = S_{\lambda}, \hspace{5mm} \lambda_i \in \mathbb{N}
\end{equation}
Super-Schur polynomials obeys the following transposing rule:
\begin{equation}
	\label{transposing rule super-Schurs}
	\mathcal{S}_{\lambda^T}(p, \theta) = \left( \cos \pi |\lambda| + \sin \pi |\lambda| \right) \cdot \mathcal{S}_{\lambda}(-p, \theta)
\end{equation}
where $|\lambda| \in \mathbb{N}/2$ and $\cos \pi |\lambda| + \sin \pi |\lambda| = \pm 1$. Using this transposing rule we can rewrite the Cauchy identity in the following form:
\begin{equation}
	\sum_{\lambda} \, \left( \cos \pi |\lambda| + \sin \pi |\lambda| \right) \cdot  \mathcal{S}_{\lambda} (p,\theta) \cdot \mathcal{S}_{\lambda^{T}}(-\bar{p},\bar{\theta}) = \exp\left( \sum_{n=1} \frac{p_n \, \bar{p}_n}{n} + \theta_n \, \bar{\theta}_n \right)
\end{equation}
\section{Super-Macdonald polynomials} \label{sec:super-Macdonald}
In this section we propose a definition of a new family of $q,t$-deformed polynomials $\mathcal{M}^{q,t}_{\lambda} (p, \theta)$ that generalizes Macdonald polynomials to the case of super-Young diagrams. We call this family super-Macdonald polynomials or Macdonald polynomials for super-Young diagrams. The definition has two main steps:

	\underline{Step 1}. We propose the ordering on the set of super-Young diagrams that generalizes the row $\prec_r$ and column $\prec_c$ ordering on the set of usual Young diagrams. We denote the generalized orderings by the same labels $\prec_{r,c}$. The main feature of the generalized ordering is that they are \underline{\textit{partial}} orderings on the set of super-Young diagrams. In particular, the orders $\prec_{r,c}$ are defined for a pair of diagrams $\lambda, \lambda^{\prime}$ only if the diagrams are of the same kind: either both contain only integers rows, either both contain at least one half-integer row. For admissible pair of diagrams the orderings $\prec_{r,c}$ are obvious generalizations of the usual lexicographical orderings defined in Section \ref{sec:Macdonald}. We provide an example from the level $4$:
	\begin{equation}
		\label{4 level example integer}
				{\begin{array}{c}
						\begin{tikzpicture}[scale=0.3]
							\foreach \i/\j in {0/0, 0/-1, 1/0, 1/-1, 2/0, 2/-1, 3/0, 3/-1}
							{
								\draw[thick] (\i,\j) -- (\i+1,\j);
							}
							\foreach \i/\j in {0/0, 1/0, 2/0, 3/0, 4/0}
							{
								\draw[thick] (\i,\j) -- (\i,\j-1);
							}
							\foreach \i/\j in {}
							{
								\draw[thick] (\i,\j) -- (\i-1,\j-1);
							}
						\end{tikzpicture}
					\end{array}} \succ_{r,c}
				{\begin{array}{c}
						\begin{tikzpicture}[scale=0.3]
							\foreach \i/\j in {0/0, 0/-1, 0/-2, 1/0, 1/-1, 2/0, 2/-1}
							{
								\draw[thick] (\i,\j) -- (\i+1,\j);
							}
							\foreach \i/\j in {0/0, 1/0, 0/-1, 1/-1, 2/0, 3/0}
							{
								\draw[thick] (\i,\j) -- (\i,\j-1);
							}
							\foreach \i/\j in {}
							{
								\draw[thick] (\i,\j) -- (\i-1,\j-1);
							}
						\end{tikzpicture}
					\end{array}} \succ_{r,c}
				{\begin{array}{c}
						\begin{tikzpicture}[scale=0.3]
							\foreach \i/\j in {0/0, 0/-1, 0/-2, 1/0, 1/-1, 1/-2}
							{
								\draw[thick] (\i,\j) -- (\i+1,\j);
							}
							\foreach \i/\j in {0/0, 1/0, 0/-1, 1/-1, 2/0, 2/-1}
							{
								\draw[thick] (\i,\j) -- (\i,\j-1);
							}
							\foreach \i/\j in {}
							{
								\draw[thick] (\i,\j) -- (\i-1,\j-1);
							}
						\end{tikzpicture}
					\end{array}} \succ_{r,c}
				{\begin{array}{c}
						\begin{tikzpicture}[scale=0.3]
							\foreach \i/\j in {0/0, 0/-1, 0/-2, 0/-3, 1/0, 1/-1}
							{
								\draw[thick] (\i,\j) -- (\i+1,\j);
							}
							\foreach \i/\j in {0/0, 1/0, 0/-1, 1/-1, 0/-2, 1/-2, 2/0}
							{
								\draw[thick] (\i,\j) -- (\i,\j-1);
							}
							\foreach \i/\j in {}
							{
								\draw[thick] (\i,\j) -- (\i-1,\j-1);
							}
						\end{tikzpicture}
					\end{array}} \succ_{r,c}
				{\begin{array}{c}
						\begin{tikzpicture}[scale=0.3]
							\foreach \i/\j in {0/0, 0/-1, 0/-2, 0/-3, 0/-4}
							{
								\draw[thick] (\i,\j) -- (\i+1,\j);
							}
							\foreach \i/\j in {0/0, 1/0, 0/-1, 1/-1, 0/-2, 1/-2, 0/-3, 1/-3}
							{
								\draw[thick] (\i,\j) -- (\i,\j-1);
							}
							\foreach \i/\j in {}
							{
								\draw[thick] (\i,\j) -- (\i-1,\j-1);
							}
						\end{tikzpicture}
					\end{array}}
	\end{equation}
	\begin{equation}
		\label{4 level example half integer}
				{\begin{array}{c}
						\begin{tikzpicture}[scale=0.3]
							\foreach \i/\j in {0/0, 0/-1, 1/0, 1/-1, 2/0, 2/-1, 3/0}
							{
								\draw[thick] (\i,\j) -- (\i+1,\j);
							}
							\foreach \i/\j in {0/0, 1/0, 2/0, 3/0, 0/-1}
							{
								\draw[thick] (\i,\j) -- (\i,\j-1);
							}
							\foreach \i/\j in {4/0, 1/-1}
							{
								\draw[thick] (\i,\j) -- (\i-1,\j-1);
							}
						\end{tikzpicture}
					\end{array}} \succ_{r,c}
				{\begin{array}{c}
						\begin{tikzpicture}[scale=0.3]
							\foreach \i/\j in {0/0, 0/-1, 0/-2, 1/0, 1/-1, 2/0}
							{
								\draw[thick] (\i,\j) -- (\i+1,\j);
							}
							\foreach \i/\j in {0/0, 1/0, 0/-1, 1/-1, 2/0}
							{
								\draw[thick] (\i,\j) -- (\i,\j-1);
							}
							\foreach \i/\j in {2/-1, 3/0}
							{
								\draw[thick] (\i,\j) -- (\i-1,\j-1);
							}
						\end{tikzpicture}
					\end{array}} \succ_{r,c}
				{\begin{array}{c}
						\begin{tikzpicture}[scale=0.3]
							\foreach \i/\j in {0/0, 0/-1, 1/0, 1/-1, 2/0, 0/-2}
							{
								\draw[thick] (\i,\j) -- (\i+1,\j);
							}
							\foreach \i/\j in {0/0, 1/0, 0/-1, 2/0, 1/-1, 0/-2}
							{
								\draw[thick] (\i,\j) -- (\i,\j-1);
							}
							\foreach \i/\j in {3/0, 1/-2}
							{
								\draw[thick] (\i,\j) -- (\i-1,\j-1);
							}
						\end{tikzpicture}
					\end{array}} \succ_{r,c}
				{\begin{array}{c}
						\begin{tikzpicture}[scale=0.3]
							\foreach \i/\j in {0/0, 0/-1, 0/-2, 1/0, 1/-1}
							{
								\draw[thick] (\i,\j) -- (\i+1,\j);
							}
							\foreach \i/\j in {0/0, 1/0, 0/-1, 1/-1, 2/0, 0/-2}
							{
								\draw[thick] (\i,\j) -- (\i,\j-1);
							}
							\foreach \i/\j in {2/-1, 1/-2}
							{
								\draw[thick] (\i,\j) -- (\i-1,\j-1);
							}
						\end{tikzpicture}
					\end{array}} \succ_{r,c}
				{\begin{array}{c}
						\begin{tikzpicture}[scale=0.3]
							\foreach \i/\j in {0/0, 0/-1, 0/-2, 0/-3, 1/0}
							{
								\draw[thick] (\i,\j) -- (\i+1,\j);
							}
							\foreach \i/\j in {0/0, 1/0, 0/-1, 1/-1, 0/-2, 1/-2, 0/-3}
							{
								\draw[thick] (\i,\j) -- (\i,\j-1);
							}
							\foreach \i/\j in {1/-3, 2/0}
							{
								\draw[thick] (\i,\j) -- (\i-1,\j-1);
							}
						\end{tikzpicture}
					\end{array}}
	\end{equation}
	Both of these two subsets are linear ordered, however there is no order between elements of \eqref{4 level example integer} and \eqref{4 level example half integer}. Being inspired by \eqref{triangularity Macdonald} we define super-Macdonald polynomials by the following triangular formula with respect to the above mentioned generalized row ordering on super-Young diagrams:
	\begin{tcolorbox}
		\begin{equation}
			\label{row triangularity super-Macdonald}
				\mathcal{M}^{q,t}_{\lambda} = \mathcal{S}_{\lambda} + \sum_{\mu \prec_r \lambda} \, \mathcal{K}_{\lambda, \mu}^{q,t} \,  \mathcal{S}_{\mu}
		\end{equation}
	\end{tcolorbox}
	This formulas fixes nearly half degrees of freedom, however the super-Macdonald-Kostka coefficients $\mathcal{K}_{\lambda, \mu}^{q,t}$ should be determined from the other reasons that we list further.
	
	\underline{Step 2}. We propose the generalization of $q,t$-scalar product for the space of polynomials of extended set of variables $(p,\theta)$:
	\begin{equation}
		\label{scalar product super-Macdonalds}
		\bra{p_{\Delta} \theta_{\Lambda}} \ket{p_{\Delta^{\prime}} \theta_{\Lambda^{\prime}}} = \delta_{\Delta, \Delta^{\prime}} \cdot \delta_{\Lambda, \Lambda^{\prime}} \cdot z_{\Delta} \cdot \prod_{k=1}^{l(\Delta)} \frac{q^{2 \Delta_k} - 1}{t^{2 \Delta_k} - 1} \cdot \boxed{\prod_{k=1}^{l(\Lambda)} q^{2 \Lambda_k}}
	\end{equation}
	We emphasized the new factors that correspond to Grassmann variables. Note that the norm of the $p_k$ variables is the same as in canonical Macdonald polynomials. We postulate the following generalized Cauchy identity corresponding to the above $q,t$-scalar product:
	\begin{tcolorbox}
		\begin{align}
			\label{Cauchy super-Macdonald}
			\sum_{\lambda} \, \frac{ \mathcal{M}^{q,t}_{\lambda}(p) \cdot \mathcal{M}^{q,t}_{\lambda}(\bar{p}) }{||\mathcal{M}^{q,t}_{\lambda}||^2 } = \exp\left( \sum_{n=1} \frac{p_n \, \bar{p}_n}{n} \, \frac{t^{2n}-1}{q^{2n}-1} + \frac{\theta_n \bar{\theta}_n}{q^{2n}} \right)
		\end{align}
	\end{tcolorbox}
	This formula completely fixes the super-Macdonald-Kostka coefficients $\mathcal{K}_{\lambda, \mu}^{q,t}$ and therefore finishes the definition of super-Macdonald polynomials. The norms $||\mathcal{M}^{q,t}_{\lambda}||$ can be also computed from the above Cauchy identity. From the definition of the generalized order and property \eqref{Schur = super-Schur} we conclude similar property for super-Macdonald polynomials for integer partitions (usual Young diagrams):
	\begin{equation}
		\mathcal{M}^{q,t}_{\lambda} = M^{q,t}_{\lambda}, \hspace{10mm} \lambda_i \in \mathbb{N}
	\end{equation}
	
	We list explicit examples of super-Macdonald polynomials from small levels:
	\begin{align}
		\begin{aligned}
			\mathcal{M}^{q,t}_{\, \varnothing} &= M^{q,t}_{\, \varnothing} = 1 \\
			\mathcal{M}^{q,t}_{\, \begin{tikzpicture}[scale=0.15]
					\foreach \i/\j in {0/0}
					{
						\draw[thick] (\i,\j) -- (\i+1,\j);
					}
					\foreach \i/\j in {0/0}
					{
						\draw[thick] (\i,\j) -- (\i,\j-1);
					}
					\foreach \i/\j in {1/0}
					{
						\draw[thick] (\i,\j) -- (\i-1,\j-1);
					}
			\end{tikzpicture}} &= \theta_1 &\hspace{10mm}
			\mathcal{M}^{q,t}_{\, \begin{tikzpicture}[scale=0.15]
					\foreach \i/\j in {0/0, 0/-1}
					{
						\draw[thick] (\i,\j) -- (\i+1,\j);
					}
					\foreach \i/\j in {0/0, 1/0}
					{
						\draw[thick] (\i,\j) -- (\i,\j-1);
					}
					\foreach \i/\j in {}
					{
						\draw[thick] (\i,\j) -- (\i-1,\j-1);
					}
			\end{tikzpicture}} &= M^{q,t}_{\, \begin{tikzpicture}[scale=0.15]
			\foreach \i/\j in {0/0, 0/-1}
			{
				\draw[thick] (\i,\j) -- (\i+1,\j);
			}
			\foreach \i/\j in {0/0, 1/0}
			{
				\draw[thick] (\i,\j) -- (\i,\j-1);
			}
			\foreach \i/\j in {}
			{
				\draw[thick] (\i,\j) -- (\i-1,\j-1);
			}
		\end{tikzpicture}} = p_1 \\
			\mathcal{M}^{q,t}_{\, \begin{tikzpicture}[scale=0.15]
					\foreach \i/\j in {0/0, 0/-1, 1/0}
					{
						\draw[thick] (\i,\j) -- (\i+1,\j);
					}
					\foreach \i/\j in {0/0, 1/0}
					{
						\draw[thick] (\i,\j) -- (\i,\j-1);
					}
					\foreach \i/\j in {2/0}
					{
						\draw[thick] (\i,\j) -- (\i-1,\j-1);
					}
			\end{tikzpicture}} &= \sqrt{2} \left( \frac{q^2 \left(t^2-1\right)}{q^2 t^2-1}p_1 \theta_1  + \frac{ q^2-1 }{q^2 t^2-1}\theta_2 \right) &\hspace{10mm}
			\mathcal{M}^{q,t}_{\, \begin{tikzpicture}[scale=0.15]
					\foreach \i/\j in {0/0, 0/-1}
					{
						\draw[thick] (\i,\j) -- (\i+1,\j);
					}
					\foreach \i/\j in {0/0, 1/0, 0/-1}
					{
						\draw[thick] (\i,\j) -- (\i,\j-1);
					}
					\foreach \i/\j in {1/-1}
					{
						\draw[thick] (\i,\j) -- (\i-1,\j-1);
					}
			\end{tikzpicture}} &= \frac{1}{\sqrt{2}} \left( p_1 \theta_1 - \theta_2 \right) &  & \\
			\mathcal{M}^{q,t}_{\, \begin{tikzpicture}[scale=0.15]
					\foreach \i/\j in {0/0, 0/-1, 1/0, 1/-1}
					{
						\draw[thick] (\i,\j) -- (\i+1,\j);
					}
					\foreach \i/\j in {0/0, 1/0, 2/0}
					{
						\draw[thick] (\i,\j) -- (\i,\j-1);
					}
					\foreach \i/\j in {}
					{
						\draw[thick] (\i,\j) -- (\i-1,\j-1);
					}
			\end{tikzpicture}} &= M^{q,t}_{\, \begin{tikzpicture}[scale=0.15]
			\foreach \i/\j in {0/0, 0/-1, 1/0, 1/-1}
			{
				\draw[thick] (\i,\j) -- (\i+1,\j);
			}
			\foreach \i/\j in {0/0, 1/0, 2/0}
			{
				\draw[thick] (\i,\j) -- (\i,\j-1);
			}
			\foreach \i/\j in {}
			{
				\draw[thick] (\i,\j) -- (\i-1,\j-1);
			}
		\end{tikzpicture}} = \frac{p_1^2}{2} + \frac{p_2}{2} &\hspace{10mm}
			\mathcal{M}^{q,t}_{\, \begin{tikzpicture}[scale=0.15]
					\foreach \i/\j in {0/0, 0/-1, 0/-2}
					{
						\draw[thick] (\i,\j) -- (\i+1,\j);
					}
					\foreach \i/\j in {0/0, 1/0, 0/-1, 1/-1}
					{
						\draw[thick] (\i,\j) -- (\i,\j-1);
					}
					\foreach \i/\j in {}
					{
						\draw[thick] (\i,\j) -- (\i-1,\j-1);
					}
			\end{tikzpicture} }  &= M^{q,t}_{\, \begin{tikzpicture}[scale=0.15]
			\foreach \i/\j in {0/0, 0/-1, 0/-2}
			{
				\draw[thick] (\i,\j) -- (\i+1,\j);
			}
			\foreach \i/\j in {0/0, 1/0, 0/-1, 1/-1}
			{
				\draw[thick] (\i,\j) -- (\i,\j-1);
			}
			\foreach \i/\j in {}
			{
				\draw[thick] (\i,\j) -- (\i-1,\j-1);
			}
		\end{tikzpicture} } = \frac{p_1^2}{2} - \frac{p_2}{2}  & & \\
			\mathcal{M}^{q,t}_{\, \begin{tikzpicture}[scale=0.15]
					\foreach \i/\j in {0/0, 0/-1, 1/0}
					{
						\draw[thick] (\i,\j) -- (\i+1,\j);
					}
					\foreach \i/\j in {0/0, 1/0, 0/-1}
					{
						\draw[thick] (\i,\j) -- (\i,\j-1);
					}
					\foreach \i/\j in {2/0, 1/-1}
					{
						\draw[thick] (\i,\j) -- (\i-1,\j-1);
			} \end{tikzpicture}} &= \theta_1 \theta_2  & & &
		\end{aligned} \nn
	\end{align}
	\begin{align}
		\begin{aligned}
			\mathcal{M}^{q,t}_{\, \begin{tikzpicture}[scale=0.15]
					\foreach \i/\j in {0/0, 0/-1, 1/0, 1/-1, 2/0, 2/-1}
					{
						\draw[thick] (\i,\j) -- (\i+1,\j);
					}
					\foreach \i/\j in {0/0, 1/0, 2/0, 3/0}
					{
						\draw[thick] (\i,\j) -- (\i,\j-1);
					}
					\foreach \i/\j in {}
					{
						\draw[thick] (\i,\j) -- (\i-1,\j-1);
					}
			\end{tikzpicture}} &= M^{q,t}_{\, \begin{tikzpicture}[scale=0.15]
					\foreach \i/\j in {0/0, 0/-1, 1/0, 1/-1, 2/0, 2/-1}
					{
						\draw[thick] (\i,\j) -- (\i+1,\j);
					}
					\foreach \i/\j in {0/0, 1/0, 2/0, 3/0}
					{
						\draw[thick] (\i,\j) -- (\i,\j-1);
					}
					\foreach \i/\j in {}
					{
						\draw[thick] (\i,\j) -- (\i-1,\j-1);
					}
			\end{tikzpicture}} & \hspace{10mm}
			\mathcal{M}^{q,t}_{\, \begin{tikzpicture}[scale=0.15]
					\foreach \i/\j in {0/0, 0/-1, 1/0, 1/-1, 2/0}
					{
						\draw[thick] (\i,\j) -- (\i+1,\j);
					}
					\foreach \i/\j in {0/0, 1/0, 2/0, 0/-1}
					{
						\draw[thick] (\i,\j) -- (\i,\j-1);
					}
					\foreach \i/\j in {3/0, 1/-1}
					{
						\draw[thick] (\i,\j) -- (\i-1,\j-1);
					}
			\end{tikzpicture}} &= \sqrt{2} \left( \frac{q^2 \left(t^2-1\right)}{q^2 t^2-1} p_1 \theta_1 \theta_2 + \frac{q^2-1}{q^2 t^2-1} \theta_1 \theta_3 \right) \\
			\mathcal{M}^{q,t}_{\,\begin{tikzpicture}[scale=0.15]
					\foreach \i/\j in {0/0, 0/-1, 0/-2, 1/0, 1/-1}
					{
						\draw[thick] (\i,\j) -- (\i+1,\j);
					}
					\foreach \i/\j in {0/0, 1/0, 0/-1, 1/-1, 2/0}
					{
						\draw[thick] (\i,\j) -- (\i,\j-1);
					}
					\foreach \i/\j in {}
					{
						\draw[thick] (\i,\j) -- (\i-1,\j-1);
					}
			\end{tikzpicture}} &= M^{q,t}_{\,\begin{tikzpicture}[scale=0.15]
					\foreach \i/\j in {0/0, 0/-1, 0/-2, 1/0, 1/-1}
					{
						\draw[thick] (\i,\j) -- (\i+1,\j);
					}
					\foreach \i/\j in {0/0, 1/0, 0/-1, 1/-1, 2/0}
					{
						\draw[thick] (\i,\j) -- (\i,\j-1);
					}
					\foreach \i/\j in {}
					{
						\draw[thick] (\i,\j) -- (\i-1,\j-1);
					}
			\end{tikzpicture}} & \hspace{10mm}
			\mathcal{M}^{q,t}_{\, \begin{tikzpicture}[scale=0.15]
					\foreach \i/\j in {0/0, 0/-1, 1/0, 0/-2}
					{
						\draw[thick] (\i,\j) -- (\i+1,\j);
					}
					\foreach \i/\j in {0/0, 1/0, 0/-1, 1/-1, 0/-2}
					{
						\draw[thick] (\i,\j) -- (\i,\j-1);
					}
					\foreach \i/\j in {2/0, 1/-2}
					{
						\draw[thick] (\i,\j) -- (\i-1,\j-1);
					}
			\end{tikzpicture}} &= \frac{1}{\sqrt{2}} \left( p_1 \theta_1 \theta_2 - \theta_1 \theta_3 \right) \\
			\mathcal{M}^{q,t}_{\, \begin{tikzpicture}[scale=0.15]
					\foreach \i/\j in {0/0, 0/-1, 0/-2, 0/-3}
					{
						\draw[thick] (\i,\j) -- (\i+1,\j);
					}
					\foreach \i/\j in {0/0, 1/0, 0/-1, 1/-1, 0/-2, 1/-2}
					{
						\draw[thick] (\i,\j) -- (\i,\j-1);
					}
					\foreach \i/\j in {}
					{
						\draw[thick] (\i,\j) -- (\i-1,\j-1);
					}
			\end{tikzpicture}} &= M^{q,t}_{\, \begin{tikzpicture}[scale=0.15]
					\foreach \i/\j in {0/0, 0/-1, 0/-2, 0/-3}
					{
						\draw[thick] (\i,\j) -- (\i+1,\j);
					}
					\foreach \i/\j in {0/0, 1/0, 0/-1, 1/-1, 0/-2, 1/-2}
					{
						\draw[thick] (\i,\j) -- (\i,\j-1);
					}
					\foreach \i/\j in {}
					{
						\draw[thick] (\i,\j) -- (\i-1,\j-1);
					}
			\end{tikzpicture}} & &
		\end{aligned}
	\end{align}
	
	\begin{align}
		\begin{aligned}
			\mathcal{M}^{q,t}_{\, \begin{tikzpicture}[scale=0.15]
					\foreach \i/\j in {0/0, 0/-1, 1/0, 1/-1, 2/0}
					{
						\draw[thick] (\i,\j) -- (\i+1,\j);
					}
					\foreach \i/\j in {0/0, 1/0, 2/0}
					{
						\draw[thick] (\i,\j) -- (\i,\j-1);
					}
					\foreach \i/\j in {3/0}
					{
						\draw[thick] (\i,\j) -- (\i-1,\j-1);
					}
			\end{tikzpicture}} &= \sqrt{3} \Biggl( \frac{q^4 \left(q^2+1\right) \left(t^2-1\right)^2}{2 \left(q^2 t^2-1\right) \left(q^4 t^2-1\right)}p_1^2 \theta_1 + \frac{ q^4 \left(q^2-1\right) \left(t^4-1\right)}{2 \left(q^2 t^2-1\right) \left(q^4 t^2-1\right)} p_2 \theta_1  +\\
			+& \frac{ q^2 \left(q^4-1\right) \left(t^2-1\right)}{\left(q^2 t^2-1\right) \left(q^4 t^2-1\right)} p_1 \theta_2 + \frac{\left(q^2-1\right)^2 \left(q^2+1\right)}{\left(q^2 t^2-1\right) \left(q^4 t^2-1\right)} \theta_3 \Biggl)\\
			\mathcal{M}^{q,t}_{\, \begin{tikzpicture}[scale=0.15]
					\foreach \i/\j in {0/0, 0/-1, 1/0, 1/-1}
					{
						\draw[thick] (\i,\j) -- (\i+1,\j);
					}
					\foreach \i/\j in {0/0, 1/0, 0/-1, 2/0}
					{
						\draw[thick] (\i,\j) -- (\i,\j-1);
					}
					\foreach \i/\j in {1/-1}
					{
						\draw[thick] (\i,\j) -- (\i-1,\j-1);
					}
			\end{tikzpicture}} &= \frac{1}{\sqrt{6}} \Biggl( \frac{\left(q^2+1\right) \left(t^2-1\right)}{ \left(q^2 t^2-1\right)}p_1^2 \theta_1 + \frac{\left(q^2-1\right) \left(t^2+1\right)}{q^2 t^2-1} p_2 \theta_1 - \frac{2 q^2 \left(t^2-1\right)}{q^2 t^2-1} p_1 \theta_2 -\frac{2 \left(q^2-1\right)}{q^2 t^2-1} \theta_3 \Biggl) \\
			\mathcal{M}^{q,t}_{\, \begin{tikzpicture}[scale=0.15]
					\foreach \i/\j in {0/0, 0/-1, 1/0, 0/-2}
					{
						\draw[thick] (\i,\j) -- (\i+1,\j);
					}
					\foreach \i/\j in {0/0, 1/0, 0/-1, 1/-1}
					{
						\draw[thick] (\i,\j) -- (\i,\j-1);
					}
					\foreach \i/\j in {2/0}
					{
						\draw[thick] (\i,\j) -- (\i-1,\j-1);
					}
			\end{tikzpicture}} &= \sqrt{\frac{3}{2}} \Biggl( \frac{q^2 \left(t^4-1\right)}{2 (q^2 t^4-1)} \left(p_1^2 \theta_1 -  p_2 \theta_1 \right) + \frac{q^2-1}{q^2 t^4-1} \left( p_1 \theta_2 - \theta_3 \right) \Biggl) \\
			\mathcal{M}^{q,t}_{\, \begin{tikzpicture}[scale=0.15]
					\foreach \i/\j in {0/0, 0/-1, 0/-2}
					{
						\draw[thick] (\i,\j) -- (\i+1,\j);
					}
					\foreach \i/\j in {0/0, 1/0, 0/-1, 1/-1, 0/-2}
					{
						\draw[thick] (\i,\j) -- (\i,\j-1);
					}
					\foreach \i/\j in {1/-2}
					{
						\draw[thick] (\i,\j) -- (\i-1,\j-1);
					}
			\end{tikzpicture} } &= \frac{1}{\sqrt{3}} \left( \frac{p_1^2 \theta_1}{2} - \frac{p_2 \theta_1}{2} - p_1 \theta_2 + \theta_3 \right) 
		\end{aligned} \nn
	\end{align}
	
	\section{Properties of super-Macdonald polynomials} \label{sec:Properties}
	
		\hspace{5mm} $\bullet$ \underline{Transposing relations}. The first interesting property that does not naively follow from the definition is the following transposing rule for super-Macdonald polynomials:
		\begin{tcolorbox}
			\begin{equation}
				\label{transposing super-Macdonald}
				\left( \cos \pi |\lambda| + \sin \pi |\lambda| \right) \cdot \frac{\mathcal{M}^{q,t}_{\lambda}(p_k,\theta_k)}{||\mathcal{M}^{q,t}_{\lambda}||^2} = \mathcal{M}^{\frac{1}{t},\frac{1}{q}}_{\lambda^{T}}\left( (-) \cdot \frac{t^{2k} - 1}{q^{2k} - 1} \cdot p_k, \frac{1}{q^{2k}} \cdot \theta_k \right)
			\end{equation}
		\end{tcolorbox}
		Note that the $q,t$ parameters should be changed according to the rule $q \to \frac{1}{t}, t \to \frac{1}{q}$. Taking into account the fact that the super-Macdonald polynomials coincide with usual Macdonald polynomials for integer partitions we obtain another transposing rule for Macdonald polynomials, because the rule \eqref{transposing rule} involves $q \leftrightarrow t$. However, there is no contradiction because the usual Macdonald polynomials have additional symmetry $q \to \frac{1}{q}, t \to \frac{1}{t}$:
		\begin{equation}
			M^{q,t}_{\lambda} = M^{\frac{1}{q},\frac{1}{t}}_{\lambda}
		\end{equation}
		The sign factor $ \cos \pi |\lambda| + \sin \pi |\lambda| = \pm 1$ is inherited directly from super-Schur polynomials \eqref{transposing rule super-Schurs}.
		Alternative form of generalized Cauchy identity, that is analog of \eqref{alternative Cauchy for Macdonald}, that reflects the transposing property:
		\begin{tcolorbox}
			\begin{align}
				\sum_{\lambda} \, \left( \cos \pi |\lambda| + \sin \pi |\lambda| \right) \cdot \mathcal{M}^{q,t}_{\lambda}(p) \cdot \mathcal{M}^{\frac{1}{t},\frac{1}{q}}_{\lambda^T}(-\bar{p})  = \exp\left( \sum_{n=1} \frac{p_n \, \bar{p}_n}{n}  + \theta_n \bar{\theta}_n \right)
			\end{align}
		\end{tcolorbox}
		$\bullet$ \underline{Triangular decompositions}. The second and the most important property in this short paper is the equivalence of row and column triangular decompositions for super-Macdonald polynomials. Note that we did not require this equivalence in the definition, only \underline{\textit{row-ordering}} is needed in \eqref{row triangularity super-Macdonald}. In other words, the following formula involving \underline{\textit{column-ordering}} is the nontrivial property of super-Macdonald polynomials:
		\begin{tcolorbox}
			\begin{equation}
				\label{column triangularity super-Macdonald}
				\mathcal{M}^{q,t}_{\lambda} = \mathcal{S}_{\lambda} + \sum_{\mu \prec_c \lambda} \, \mathcal{K}_{\lambda, \mu}^{q,t} \,  \mathcal{S}_{\mu}
			\end{equation}
		\end{tcolorbox}
		Row and column ordering on the set of super-Young diagrams starts to differ on $\frac{9}{2}$ level:
		\begin{equation}
				\begin{array}{c}
					\begin{tikzpicture}[scale=0.3]
					\foreach \i/\j in {0/0, 0/-1, 0/-2, 1/0, 1/-1, 2/0, 2/-1}
					{
						\draw[thick] (\i,\j) -- (\i+1,\j);
					}
					\foreach \i/\j in {0/0, 1/0, 0/-1, 1/-1, 2/0, 3/0, 0/-2}
					{
						\draw[thick] (\i,\j) -- (\i,\j-1);
					}
					\foreach \i/\j in {1/-2}
					{
						\draw[thick] (\i,\j) -- (\i-1,\j-1);
					}
				\end{tikzpicture}
			\end{array} \succ_{r}
			\begin{array}{c}
				\begin{tikzpicture}[scale=0.3]
					\foreach \i/\j in {0/0, 0/-1, 0/-2, 1/0, 1/-1, 2/0, 1/-2}
					{
						\draw[thick] (\i,\j) -- (\i+1,\j);
					}
					\foreach \i/\j in {0/0, 1/0, 0/-1, 1/-1, 2/0, 2/-1}
					{
						\draw[thick] (\i,\j) -- (\i,\j-1);
					}
					\foreach \i/\j in {3/0}
					{
						\draw[thick] (\i,\j) -- (\i-1,\j-1);
					}
				\end{tikzpicture}
			\end{array} \hspace{15mm}
			\begin{array}{c}
				\begin{tikzpicture}[scale=0.3]
					\foreach \i/\j in {0/0, 0/-1, 0/-2, 1/0, 1/-1, 2/0, 2/-1}
					{
						\draw[thick] (\i,\j) -- (\i+1,\j);
					}
					\foreach \i/\j in {0/0, 1/0, 0/-1, 1/-1, 2/0, 3/0, 0/-2}
					{
						\draw[thick] (\i,\j) -- (\i,\j-1);
					}
					\foreach \i/\j in {1/-2}
					{
						\draw[thick] (\i,\j) -- (\i-1,\j-1);
					}
				\end{tikzpicture}
			\end{array}
			 \prec_{c}
			\begin{array}{c}
				\begin{tikzpicture}[scale=0.3]
					\foreach \i/\j in {0/0, 0/-1, 0/-2, 1/0, 1/-1, 2/0, 1/-2}
					{
						\draw[thick] (\i,\j) -- (\i+1,\j);
					}
					\foreach \i/\j in {0/0, 1/0, 0/-1, 1/-1, 2/0, 2/-1}
					{
						\draw[thick] (\i,\j) -- (\i,\j-1);
					}
					\foreach \i/\j in {3/0}
					{
						\draw[thick] (\i,\j) -- (\i-1,\j-1);
					}
				\end{tikzpicture}
			\end{array}
		\end{equation}
		
		\begin{equation}
			\begin{array}{c}
				\begin{tikzpicture}[scale=0.3]
					\foreach \i/\j in {0/0, 0/-1, 0/-2, 1/0, 1/-1, 2/0, 0/-3}
					{
						\draw[thick] (\i,\j) -- (\i+1,\j);
					}
					\foreach \i/\j in {0/0, 1/0, 0/-1, 1/-1, 2/0, 0/-2, 1/-2}
					{
						\draw[thick] (\i,\j) -- (\i,\j-1);
					}
					\foreach \i/\j in {3/0}
					{
						\draw[thick] (\i,\j) -- (\i-1,\j-1);
					}
				\end{tikzpicture}
			\end{array} \succ_{r}
			\begin{array}{c}
				\begin{tikzpicture}[scale=0.3]
					\foreach \i/\j in {0/0, 0/-1, 0/-2, 1/0, 1/-1, 1/-2}
					{
						\draw[thick] (\i,\j) -- (\i+1,\j);
					}
					\foreach \i/\j in {0/0, 1/0, 0/-1, 1/-1, 2/0, 2/-1, 0/-2}
					{
						\draw[thick] (\i,\j) -- (\i,\j-1);
					}
					\foreach \i/\j in {1/-2}
					{
						\draw[thick] (\i,\j) -- (\i-1,\j-1);
					}
				\end{tikzpicture}
			\end{array} \hspace{15mm}
			\begin{array}{c}
				\begin{tikzpicture}[scale=0.3]
					\foreach \i/\j in {0/0, 0/-1, 0/-2, 1/0, 1/-1, 2/0, 0/-3}
					{
						\draw[thick] (\i,\j) -- (\i+1,\j);
					}
					\foreach \i/\j in {0/0, 1/0, 0/-1, 1/-1, 2/0, 0/-2, 1/-2}
					{
						\draw[thick] (\i,\j) -- (\i,\j-1);
					}
					\foreach \i/\j in {3/0}
					{
						\draw[thick] (\i,\j) -- (\i-1,\j-1);
					}
				\end{tikzpicture}
			\end{array} \prec_{c}
			\begin{array}{c}
				\begin{tikzpicture}[scale=0.3]
					\foreach \i/\j in {0/0, 0/-1, 0/-2, 1/0, 1/-1, 1/-2}
					{
						\draw[thick] (\i,\j) -- (\i+1,\j);
					}
					\foreach \i/\j in {0/0, 1/0, 0/-1, 1/-1, 2/0, 2/-1, 0/-2}
					{
						\draw[thick] (\i,\j) -- (\i,\j-1);
					}
					\foreach \i/\j in {1/-2}
					{
						\draw[thick] (\i,\j) -- (\i-1,\j-1);
					}
				\end{tikzpicture}
			\end{array}
		\end{equation}
		However, super-Macdonald polynomials respect both orderings and therefore the super-Macdonald-Kostka coefficients corresponding to the above diagrams vanish:
		\begin{align}
			\begin{aligned}
				\mathcal{K}^{q,t}_{\left[ 3,1,\frac{1}{2} \right], \left[ \frac{5}{2},2 \right]} &= 0 &\hspace{15mm} \mathcal{K}^{q,t}_{\left[ \frac{5}{2},2 \right], \left[ 3,1,\frac{1}{2} \right]} &= 0 \\
				\mathcal{K}^{q,t}_{\left[ \frac{5}{2}, 1, 1 \right], \left[ 2,2, \frac{1}{2} \right]} &= 0 &\hspace{15mm} \mathcal{K}^{q,t}_{\left[ 2,2, \frac{1}{2} \right], \left[ \frac{5}{2}, 1, 1 \right]} &= 0 \\
			\end{aligned}
		\end{align}
		We do not list more explicit examples here, instead we give the following statistics. We consider diagrams $\lambda$ and $\lambda^{\prime}$ with the condition:
		\begin{equation}
			\lambda \succ_{r} \lambda^{\prime} \hspace{5mm} \text{but} \hspace{5mm} \lambda \prec_{c} \lambda^{\prime}
		\end{equation}
		The numbers of such diagrams for integer partitions (usual Young diagrams) and half-integer partitions (super-Young diagrams containing at least one half-integer row) are collected in the Table \ref{Stats}:
		\begin{table}[h!]
			\centering
			$\begin{array}{|c|c|c|c|c|c|c|c|c|c|c|c|c|c|}
				\hline
				\text{Level} & 9/2 & 5 & 11/2 & 6 & 13/2 & 7 & 15/2 & 8 & 17/2 & 9 & 19/2 & 10 & 21/2\\
				\hline
				\text{Half-integer partitions} & 2 & 0 & 10 & 8 & 48 & 27 & 146 & 115 & 435 & 331 & 1151 & 993 & 2953 \\
				\hline
				\text{Integer partitions} & 0 & 0 & 0 & 2 & 0 & 4 & 0 & 15 & 0 & 35 & 0 & 85 & 0 \\
				\hline
			\end{array}$
			\caption{The numbers of pairs of diagrams where two orderings are different.}
			\label{Stats}
		\end{table}
	
		For all the above cases of diagrams $\lambda$, $\lambda^{\prime}$ on the levels up to $21/2$ we checked by calculation that the corresponding super-Macdonald-Kostka coefficients vanish:
		\begin{equation}
			\mathcal{K}^{q,t}_{\lambda, \lambda^{\prime}} = 0 \hspace{10mm}  \mathcal{K}^{q,t}_{\lambda^{\prime}, \lambda} = 0
		\end{equation}
		Note that the vanishing of the corresponding super-Macdonald-Kostka coefficients in case of integer partitions follows directly from the analogous property of usual Macdonald polynomials, because in this case they are equal to usual Macdonald-Kostka coefficients:
		\begin{equation}
			\mathcal{K}^{q,t}_{\lambda, \lambda^{\prime}} = K^{q,t}_{\lambda, \lambda^{\prime}} \hspace{10mm} \lambda_i, \lambda^{\prime}_i \in \mathbb{N}
		\end{equation}
		We elaborate more on the peculiarities of this triangular (Kerov) approach to super-Macdonals in a separate paper \cite{GMT2}. Like in the case of ordinary Macdonalds the resulting polynomials
		do not depend on the ordering (what is not the case for generic Kerov functions), but unlike that case the Kostka coefficients will not be fully factorizable.
		
		$\bullet$ \underline{Limit of super-Macdonald polynomials}. The limit to super-Schur polynomials has two steps: 1) set $t = q$ and then 2) take a limit $q \to 1$. Note that usual Macdonald polynomials becomes Schur polynomials already on the first step $t = q$.
		
		$\bullet$ \underline{Hamiltonians for super-Macdonald polynomials}. The residue formula for Macdonald Hamiltonian \eqref{Macdonald H} can be easily represented in terms of Schur polynomials:
		\begin{equation}
			\label{Mac Ham via Schur}
			\hat{H} = \sum_{n = 0} S_{[n]} (p_k^{*}) \cdot S_{[n]} ( \hat{p}_k )
		\end{equation}
		where we introduced new notations for rescaled variables:
		\begin{equation}
			\begin{aligned}
				p_k^{*} := \left( 1 - t^{-2k} \right) p_k \hspace{25mm} \hat{p}_k := \left( q^{2k} - 1 \right)\cdot k \frac{\partial}{\partial p_k} 
			\end{aligned}
		\end{equation}
		Miraculously, the dependence on $q,t$ parameters in the above Hamiltonian \eqref{Mac Ham via Schur} is completely fixed by these rescaling rules. Recall that the eigenvalues have the following form:
	\begin{equation}
		\mathcal{E}^{\, q,t}_{\lambda} = 1 + \gamma \cdot \sum_{\Box \in \lambda} q^{2 j_{\Box}} t^{-2 i_{\Box}} \hspace{20mm} \gamma := (q^2 + 1)(1 - t^{-2})
	\end{equation}
	Using these eigenvalues as the base point we define a new Hamiltonian $\hat{\mathcal{H}}^{+}$ for super-Macdonald polynomials by the following rules:
	\begin{align}
		\hat{\mathcal{H}}^{+} \, \mathcal{M}^{q,t}_{\lambda} &= \mathcal{E}^{+,\, q,t}_{\lambda} \, \mathcal{M}^{q,t}_{\lambda} \hspace{25mm}
		\mathcal{E}^{+,\, q,t}_{\lambda} = 1 + \gamma \cdot \sum_{\, \begin{tikzpicture}[scale=0.15]
				\foreach \i/\j in {0/0}
				{
					\draw[thick] (\i,\j) -- (\i+1,\j);
				}
				\foreach \i/\j in {0/0}
				{
					\draw[thick] (\i,\j) -- (\i,\j-1);
				}
				\foreach \i/\j in {1/0}
				{
					\draw[thick] (\i,\j) -- (\i-1,\j-1);
				}
			\end{tikzpicture} \in \lambda} q^{2 j_{\, \begin{tikzpicture}[scale=0.15]
					\foreach \i/\j in {0/0}
					{
						\draw[thick] (\i,\j) -- (\i+1,\j);
					}
					\foreach \i/\j in {0/0}
					{
						\draw[thick] (\i,\j) -- (\i,\j-1);
					}
					\foreach \i/\j in {1/0}
					{
						\draw[thick] (\i,\j) -- (\i-1,\j-1);
					}
		\end{tikzpicture}}} t^{-2 i_{\, \begin{tikzpicture}[scale=0.15]
					\foreach \i/\j in {0/0}
					{
						\draw[thick] (\i,\j) -- (\i+1,\j);
					}
					\foreach \i/\j in {0/0}
					{
						\draw[thick] (\i,\j) -- (\i,\j-1);
					}
					\foreach \i/\j in {1/0}
					{
						\draw[thick] (\i,\j) -- (\i-1,\j-1);
					}
		\end{tikzpicture}}} 
	\end{align}
	In the above formula for eigenvalues $\mathcal{E}^{+,\, q,t}_{\lambda}$ the sum goes over all half-boxes $\begin{tikzpicture}[scale=0.15]
		\foreach \i/\j in {0/0}
		{
			\draw[thick] (\i,\j) -- (\i+1,\j);
		}
		\foreach \i/\j in {0/0}
		{
			\draw[thick] (\i,\j) -- (\i,\j-1);
		}
		\foreach \i/\j in {1/0}
		{
			\draw[thick] (\i,\j) -- (\i-1,\j-1);
		}
	\end{tikzpicture}$ including those that enter the full boxes. These new eigenvalues $\mathcal{E}^{+,\, q,t}_{\lambda}$ coincide with canonical ones $\mathcal{E}^{\, q,t}_{\lambda}$ for integer partitions (without half-integers). We computed the first several orders of the new super-Hamiltonian:
	\begin{align} 
		\begin{aligned}
			\mathcal{H}^{+} &= 1 + \theta^*_1\hat\theta_1 + \overbrace{S_{[1]}(p^*_k)S_{[1]}(\hat p_k)}^{S_{[1]}^*\hat S_{[1]}} + \\
			&+ \left\{ (S_{[1]}^*\theta_1^* + \theta^*_2)(\hat S_{[1]}\hat\theta_1  + \hat\theta_2) + \left(\frac{t}{q}\right)^2\theta^*_2\hat\theta_2 \right\}
			+ \left\{S_{[2]}^*\hat S_{[2]} + \left(\frac{t}{q}\right)^2\theta^*_1\theta^*_2\hat\theta_1 \hat\theta_2 \right\} + \\
			&+\left\{ ( S_{[2]}^* \theta_1^* + S_{[1]}^* \theta_2^{*} + \theta_3^{*} )( \hat S_{[2]} \hat\theta_1 + \hat S_{[1]} \hat\theta_2 + \hat\theta_3 ) + \left(\frac{t}{q}\right)^2 (S_{[1]}^*\theta_2^* + \theta^*_3)(\hat S_{[1]}\hat\theta_2  + \hat\theta_3)  + \left(\frac{t}{q}\right)^4 \theta^*_3\hat\theta_3\right\}  + \ldots
		\end{aligned}
		\label{super Hamiltonian}
	\end{align}
	In this case the $q,t$-dependence is almost completely fixed provided the following rescaling rules for Grassmann varibles are satisfied:
	\begin{equation}
		\begin{aligned}
			\theta_k^{*} := t^{-2(k-1)} \left( 1 - t^{-2}\right) \theta_k \hspace{25mm}
			 \hat{\theta}_k := q^{2(k-1)} \left( q^{2} - 1 \right) \frac{\partial}{\partial \theta_k}
		\end{aligned}
	\end{equation}
	We conjecture that the remaining $q,t$-dependence enter only by $(t/q)^{2n}$ factors. The complete formula for super-Hamiltonian $\mathcal{H}^{+}$ in the form like \eqref{Macdonald H} requires additional notation and comments. It will be presented in a separate publication \cite{GMT3}.

	\section{Conclusion} \label{sec:Conclusion}
	In this short paper we introduced the generalization of Macdonald polynomials for super-Young diagrams that may contain half-boxes. Our definition is based on super-Schur polynomials that naturally arise as the space of semi-Fock representation of affine super-Yangian $Y(\hat{\mathfrak{gl}}_{1|1})$ \cite{Galakhov:2023mak}. Starting from orthogonality relations \eqref{Cauchy Macdonald} and triangular decompositions \eqref{row triangularity Macdonald} of classical Macdonald polynomials we define super-Macdonald polynomials postulating similar properties for the super-case: \eqref{Cauchy super-Macdonald} and \eqref{row triangularity super-Macdonald} respectively. 
	
	We perform several obvious checks to confirm the reasonableness of the new super-Macdonald polynomials. Turns out that properties of canonical Macdonald polynomials such as transposing relations \eqref{transposing rule} and existence of Hamiltonians \eqref{Macdonald H} are lifted to the level of super-ones, see \eqref{transposing super-Macdonald} and \eqref{super Hamiltonian}. Among these properties we distinguish the following: \textit{super-Macdonald polynomials respect two natural orderings \eqref{row triangularity super-Macdonald} and \eqref{column triangularity super-Macdonald} on the set of super-Young diagrams}. This property is a direct generalization of the corresponding property of Macdonald polynomials presented in \cite{Mironov:2020aaa}. In \cite{Mironov:2020aaa} it was shown in detail that this respect of two orders is a special property of Macdonald measure \eqref{q,t measure Macdonald}. In particular, the respect of two orders vanishes for the more general Kerov polynomials \cite{kerov1991hall}. We check the respect of two orders property for the super-Macdonald polynomials by explicit calculation up to 21/2 level. We consider this fact as a rather strong argument in favour of our $(q,t)$-measure \eqref{scalar product super-Macdonalds}.
	
	Super-Macdonald polynomials are $(q,t)$-deformation of super-Schur polynomials therefore they should form a representation of $(q,t)$-deformation of affine super-Yangian $Y(\hat{\mathfrak{gl}}_{1|1})$ that is called toroidal algebra $T(\hat{\mathfrak{gl}}_{1|1})$ or super-DIM algebra \cite{Noshita:2021dgj, Galakhov:2021vbo} that is a close relative of usual DIM algebra/toroidal algebra $T(\hat{\mathfrak{gl}}_{1})$ \cite{Ding:1996mq, Miki:2007mer, Awata:2017lqa, Feigin2}. Another perspective direction is a question about commuting set of super-Hamiltonians that define super-Macdonald polynomials as a common set of eigenfunctions. These super-Hamiltonians will form a Cartan subalgebra of the toroidal algebra $T(\hat{\mathfrak{gl}}_{1|1})$. The analysis of these questions will be presented in a separate paper \cite{GMT3}. 
	
	Hope this short paper will gain an interest and encourage researchers to prove the above mentioned and find novel interesting properties of super-Macdonald polynomials.
	\section*{Acknowledgments}
	Our work is supported by the Russian Science Foundation (Grant No. 20-71-10073).

	\bibliographystyle{utphys}
	\bibliography{biblio}
\end{document}